\documentclass[sigconf]{acmart}
\usepackage{array}

\AtBeginDocument{%
  }

\copyrightyear{2025} 
\acmYear{2025} 
\setcopyright{acmlicensed}\acmConference[UIST '25]{The 38th Annual ACM Symposium on User Interface Software and Technology}{September 28-October 1, 2025}{Busan, Republic of Korea}
\acmBooktitle{The 38th Annual ACM Symposium on User Interface Software and Technology (UIST '25), September 28-October 1, 2025, Busan, Republic of Korea}
\acmDOI{10.1145/3746059.3747793}
\acmISBN{979-8-4007-2037-6/2025/09}

\begin{document}

\title{CineVision: An Interactive Pre-visualization Storyboard System for Director–Cinematographer Collaboration}

\author{Zheng Wei}
\authornote{Both authors contributed equally to this research.}
\email{zwei302@connect.ust.hk}
\affiliation{%
  \institution{The Hong Kong University of Science and Technology}
  \city{Hong Kong}
  \country{China}
}

\author{Hongtao Wu}
\authornotemark[1]
\email{wuhongtao@westlake.edu.cn}
\affiliation{%
  \institution{The Hong Kong University of Science and Technology $\&$ Westlake University}
  \city{Hong Kong}
  \country{China}
  }

\author{Lvmin Zhang}
\email{lvmin@cs.stanford.edu}
\affiliation{%
  \institution{Stanford University}
  \city{Palo Alto}
  \country{USA}
}

\author{Xian Xu}
\email{xianxu@ust.hk}\email{xianxu0523@gmail.com}
\affiliation{%
  \institution{The Hong Kong University of Science and Technology $\&$ Lingnan University}
  \city{Hong Kong}
  \country{China}
}

\author{Yefeng Zheng}
\email{zhengyefeng@westlake.edu.cn}
\affiliation{%
  \institution{Westlake University}
  \city{Hang Zhou}
  \country{China}}

\author{Pan Hui}
\email{panhui@ust.hk}
\affiliation{%
  \institution{The Hong Kong University of Science and Technology}
  \city{Hong Kong}
  \country{China}}

\author{Maneesh Agrawala}
\email{maneesh@cs.stanford.edu}
\affiliation{%
  \institution{Stanford University}
  \city{Palo Alto}
  \country{USA}
}

\author{Huamin Qu}
\email{huamin@cse.ust.hk}
\affiliation{%
  \institution{The Hong Kong University of Science and Technology}
  \city{Hong Kong}
  \country{China}
}

\author{Anyi Rao}
\email{anyirao@ust.hk}
\authornote{Corresponding author}
\affiliation{%
  \institution{The Hong Kong University of Science and Technology}
  \city{Hong Kong}
  \country{China}
}

\renewcommand{\shortauthors}{Wei et al.}

\begin{abstract}
Effective communication between directors and cinematographers is fundamental in film production, yet traditional approaches relying on visual references and hand-drawn storyboards often lack the efficiency and precision necessary during pre-production. We present \textit{CineVision}, an AI-driven platform that integrates scriptwriting with real-time visual pre-visualization to bridge this communication gap. By offering dynamic lighting control, style emulation based on renowned filmmakers, and customizable character design, \textit{CineVision} enables directors to convey their creative vision with heightened clarity and rapidly iterate on scene composition. {In a 24-participant lab study, \textit{CineVision} yielded shorter task times and higher usability ratings than two baseline methods, suggesting a potential to ease early-stage communication and accelerate storyboard drafts under controlled conditions.} These findings underscore \textit{CineVision’s} potential to streamline pre-production processes and foster deeper creative synergy among filmmaking teams, particularly for new collaborators. {Our code and demo are available at https://github.com/TonyHongtaoWu/CineVision.}
\end{abstract}



\begin{teaserfigure}
  \includegraphics[width=\textwidth]{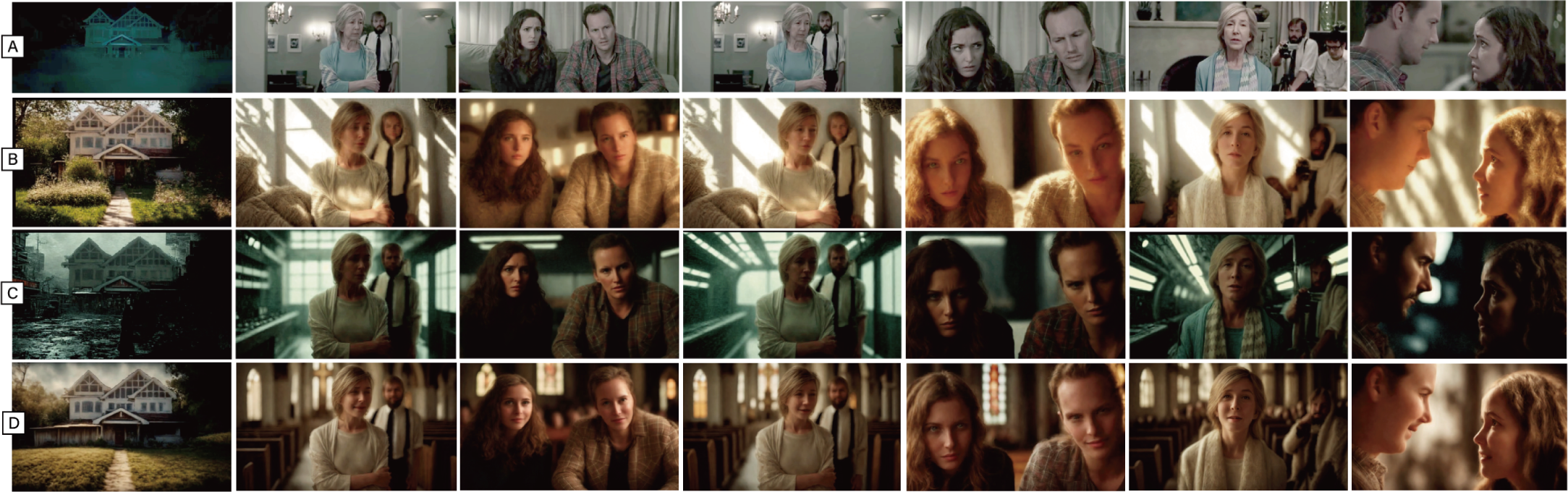}
  \caption{The images showcase a progression of modifications made using the \textit{CineVision} system. Group A features scenes from Insidious (2010), a U.S. horror film directed by James Wan, where a couple's son falls into a coma and his soul is trapped in a supernatural dimension. Group B shows how the user applies the \textit{CineVision} system to alter various aspects of the film, including background, lighting, shooting time, facial features, and actor costumes, offering a customized reinterpretation of the original visuals. Group C demonstrates the use of the ``Ridley Scott Director Style'' feature within \textit{CineVision}, reimagining Insidious with Ridley Scott’s signature style, influencing lighting, composition, and overall atmosphere. Finally, Group D emphasizes a shift in emotional tone, transforming the scene from somber to uplifting with adjustments to facial expressions, such as bright eyes and a wide smile, set against a church backdrop.}
  \Description{The images showcase a progression of modifications made using the \textit{CineVision} system. Group A features scenes from Insidious (2010), a U.S. horror film directed by James Wan, where a couple's son falls into a coma and his soul is trapped in a supernatural dimension. Group B shows how the user applies the \textit{CineVision} system to alter various aspects of the film, including background, lighting, shooting time, facial features, and actor costumes, offering a customized reinterpretation of the original visuals. Group C demonstrates the use of the ``Ridley Scott Director Style'' feature within \textit{CineVision}, reimagining Insidious with Ridley Scott’s signature style, influencing lighting, composition, and overall atmosphere. Finally, Group D emphasizes a shift in emotional tone, transforming the scene from somber to uplifting with adjustments to facial expressions, such as bright eyes and a wide smile, set against a church backdrop.}
  \label{fig:teaser}
\end{teaserfigure}

\maketitle

\section{INTRODUCTION}

Film production is a complex process that transforms a script into an immersive cinematic experience \cite{mateer2017directing,prince2011digital}. 
 
The goal of scripting is to fully convey essential visual elements such as lighting, camera angles, character positioning, and atmosphere, which are crucial for realizing the director's creative vision \cite{brewster2011fundamentals}. 
{Directors rely on} mental visualizations to conceptualize scenes~\cite{lobrutto2002filmmaker}. Translating these ideas into a common language for effective communication with production departments - such as cinematography, art direction, and lighting - remains a significant challenge~\cite{lobrutto2002filmmaker,brown2016cinematography}. 
Storyboards are typically used in these processes, while conveying detailed visual components with efficient communication remains an open problem \cite{simon2012storyboards}.
Typical inefficiencies are delays and a final product that deviates from the director's original vision \cite{katz1991film}.

Emerging platforms, such as \textit{Scriptviz} \cite{rao2024scriptviz} and \textit{VDS}~\cite{rao2023dynamic} have discussed the storyboard processes, with \textit{Scriptviz} helping screenwriters refine scripts and \textit{VDS} focusing on storyboards in a 3D game engine. However, \textit{Scriptviz} only searches existing film stills based on script descriptions and does not adjust visual elements like lighting or character design. Similarly, \textit{VDS} is constrained by its 3D game engine environment, which does not align with film production and lacks support for adjusting key visual elements. These limitations require separate, manual adjustments, resulting in fragmented workflows that reduce the efficiency of filmmaking.

To address these challenges, we introduce \textit{CineVision}, a platform designed to seamlessly integrate scriptwriting with visual pre-visualization, providing directors with a dynamic and efficient storyboard creation system that allows them to visualize and refine scenes in real time as they develop the script. \textit{CineVision} utilizes an extensive film database to match script descriptions with accurate, contextually relevant visuals, capturing the spatial and emotional dynamics of each scene. This bridges the gap between the written narrative and the director’s visual concepts, enabling directors to see their scripts come to life. It also provides rich reference material to facilitate collaboration with key production teams, such as cinematographers. 
Unlike previous methods, \textit{CineVision} integrates image re-lighting, style control, and customizable character design directly into the pre-visualization workflow. This allows directors to dynamically adjust lighting—modifying intensity, direction, and color—to evoke specific emotions and enhance the emotional tone of a scene \cite{zhang2024ic,zhang2023adding}. The system also applies the visual art styles of master directors, providing instant visual feedback on how style choices impact the overall look. Additionally, \textit{CineVision} enables real-time customization of character designs, allowing directors to modify facial expressions, hairstyles, costumes, and accessories in sync with the evolving script. This integration ensures that all visual elements, from lighting to character design, align with the director's vision throughout pre-production.

In a user study involving 24 participants, \textit{CineVision} {demonstrated its potential to boost collaboration efficiency between directors and cinematographers,} allowing them to focus on deeper narrative aspects such as character emotions and relationships, rather than technical adjustments. Compared to traditional hand-drawn storyboards and AI-assisted tools, \textit{CineVision} reduced physical and temporal demands and was rated higher in usefulness and ease of use. These findings underscore \textit{CineVision}’s ability to simplify pre-visualization, foster more nuanced collaboration, and improve overall production efficiency. By providing a seamless, interactive environment, \textit{CineVision} enhances creative control, streamlines communication between departments, and reduces the need for repeated revisions. Our work makes two main contributions: 

(1) Integrating scriptwriting with visual pre-visualization to strengthen director-cinematographer collaboration and streamline storyboard creation.

(2) Empowering directors to express their creative vision more precisely through AI-driven dynamic lighting, customizable character design, and context-sensitive visualization features.

\section{BACKGROUND AND RELATED WORK}

\subsection{Story Visualization and Storyboard}
Storyboards are essential in film and media production, providing a visual representation of scenes and shots before filming. They serve as a critical communication tool between directors and key departments like cinematography, art, sound, costume design, and acting \cite{hart2013art,block2020visual, bartindale2012storycrate,wei2024hearing}. Traditionally, storyboards have been created by manually drawing each keyframe, a process that is both time-consuming and limited by the artist’s skills and the production timeline. While traditional storyboards can be interactive, they still face constraints in terms of flexibility and detail. Recent advancements in automation, such as Generative Adversarial Networks (GANs) and Variational Autoencoders (VAEs), have streamlined this process \cite{zhang2025generative}. These generative techniques allow for the creation of corresponding visual scenes from scripts with improved realism and consistency, addressing some of the limitations of traditional methods \cite{liu2021generative,bengesi2024advancements,hong2023visual}. Large-scale datasets, such as the Visual Writing Prompts (VWP) dataset \cite{hong2023visual}, have become valuable resources for training character-based visual narrative models, further advancing automatic storyboard generation.

The development of intelligent writing assistants has also accelerated this progress. These tools can generate text proposals from keywords \cite{fan2018hierarchical,ippolito2019unsupervised,xu2020megatron,xu2024skipwriter}, engage in interactive conversations \cite{yuan2022wordcraft}, or process multimodal inputs \cite{li2022blip}, thereby enhancing the creative process. These systems have shown their effectiveness in supporting story writing \cite{biermann2022tool, clark2018creative,mirowski2023co,li2024words}, with many new intelligent writing tools emerging \cite{zhao2023leveraging,rao2023dynamic}. For example, the \textit{ScriptViz} developed by Rao et al. \cite{rao2024scriptviz} focuses on script analysis and image retrieval, matching textual descriptions with pre-existing images. However, these methods rely on predefined templates, limiting flexibility in generating novel visuals. As machine learning and computer vision have advanced \cite{croitoru2023diffusion}, methods such as Stable Diffusion have shown immense potential in automatically generating high-quality, contextually accurate visuals from text descriptions \cite{bengesi2024advancements,liu2023instaflow}. For instance, a director could describe a scene as ``a car chasing down a wet city street at night,'' and the model would generate images that accurately reflect the lighting \cite{yin2023cle}, emotions \cite{yang2024emogen}, and composition of the scene \cite{zhang2024brush}, offering greater creative flexibility.

These models have the capability to render intricate details, such as lighting effects \cite{zeng2024dilightnet} and complex camera angles \cite{he2024cameractrl}, which were previously difficult to achieve manually. The integration of style transfer techniques also allows directors to apply specific environment styles (lighting, color) to generated scenes, enhancing their creative control. However, challenges remain in maintaining stylistic consistency across scenes and adapting to the diverse preferences of directors \cite{pal2025illuminating}. Additionally, integrating complex narrative structures into continuous storyboard shots introduces further challenges, such as ensuring continuity in actors' movements, background consistency, and smooth transitions in lighting. To address these issues, our system offers a personalized, director-centric solution. By enabling directors to define visual preferences and control stylistic elements within continuous storyboard shots, our interactive AI storyboard enhances the pre-visualization process with features like re-lighting, style control, and customizable character design, providing greater creative flexibility and control.

\subsection{Image Relighting and Cinematic Lighting}

Lighting is a fundamental aspect of filmmaking, shaping both the mood and visual aesthetic of a scene \cite{keating2009hollywood,brown2016cinematography}. It has long been an essential tool for cinematographers, enabling them to evoke specific emotions and emphasize key elements within a scene {\cite{wei2023feeling,xu2024transforming,wei2025illuminating,wei2024multi}}. At the same time, image relighting—the process of altering the lighting conditions in an image—has emerged as a crucial technique in computer vision and computational photography \cite{pandey2021total,hou2021towards}. This method allows for the manipulation of lighting in digital images, offering new opportunities for visual experimentation and enhancing realism in image processing \cite{li2022physically}.

While traditional image relighting methods involved manual adjustments through photo-editing software, these processes were often slow and required significant expertise \cite{concepcion2022adobe}. As a result, there has been a growing interest in automating image relighting using machine learning techniques \cite{pandey2021total}. Recent developments in deep learning have significantly advanced the automation of image relighting, allowing for realistic and dynamic lighting effects \cite{zhang2024ic}. Techniques such as Transformers and Diffusion Models have proven effective in simulating complex lighting conditions, including changes in light intensity, direction, and color \cite{ponglertnapakorn2023difareli,yue2023dif}. These models learn from large datasets of real-world lighting scenarios, enabling them to generate plausible lighting effects that align with the physical properties of light. Moreover, the integration of physics-based rendering models with machine learning algorithms has further enhanced the physical accuracy of relighting techniques, ensuring both visual appeal and realism. In the context of film production, image relighting plays a crucial role in pre-visualization, enabling directors to experiment with different lighting \cite{wei2023feeling} setups before shooting \cite{hart2013art,glebas2012directing}. For example, systems like IC-Light leverage real-time relighting based on scene metadata, allowing creators to visualize how various lighting configurations will impact a shot~\cite{zhang2024ic}.

Our system integrates image relighting technology into the storyboard generation process, offering a level of detail that supports the director-cinematographer collaboration. By allowing directors to specify lighting parameters during pre-production, the system ensures that generated visualizations align with their artistic vision. 

\subsection{Design Character and Costume Creation}
Character design, including clothing, makeup, and facial expressions, plays a pivotal role in the pre-production process, as it helps bring the characters of the narrative to life \cite{sciortino2020pre,melki2019investigation}. These visual elements are essential in conveying the director’s vision and guiding the cinematographer in capturing the right tone and emotion, ensuring consistency and coherence throughout the film. Traditionally, creating detailed character designs for storyboards has been a manual and {labor-intensive} process, which requires skilled artists to visualize and draw characters in specific settings and costumes \cite{doust2023laboratory}. This approach is not only time-consuming, but is also often limited by the artist’s ability to translate textual descriptions into visual representations. Recent advancements in AI and machine learning, however, are transforming this process by enabling the automatic generation of highly detailed character designs and costume variations, based on textual descriptions or even sketches ~\cite{guo2023ai,he2024dresscode}.

AI-driven models~\cite{wu2023mask,wu2024rainmamba, wu2024semi, lu2023tf, lu2024mace,lu2024robust, yang2024genuine} like GANs and diffusion models have demonstrated significant capabilities in generating realistic human figures, clothing designs, and makeup styles \cite{painguzhali2025artificial,zhang2024unlocking}. These models have been trained on large datasets of fashion, historical clothing, and character designs, allowing them to create diverse and contextually appropriate visual representations of characters, from modern fashion to fantasy attire \cite{hosseini2024generative}. Moreover, the integration of AI technologies for the generation of makeup and facial characteristics is gaining traction \cite{chen2023survey}. Diffusion models can adjust facial attributes such as age, gender, expression, and even specific cosmetic styles (e.g., makeup for a period drama or a fantasy setting) \cite{boutros2023idiff,kim2023dcface,ding2023diffusionrig}. This allows for dynamic customization of characters, enabling directors to explore different visual interpretations of their characters quickly. A key area of research in this field is the generation of multi-dimensional character traits that go beyond just clothing and makeup. For example, diffusion models now allow for detailed customizations of facial features, hairstyles, accessories, and even the textures and materials of clothing \cite{samy2025revolutionizing,kim2024gala,gal2023encoder}. AI can modify these attributes interactively, providing a flexible and creative space for filmmakers to refine their character designs during pre-production.

Our system leverages these advancements in diffusion models by offering a multi-dimensional approach to character and costume creation. Directors can specify parameters such as facial features, hairstyles, clothing textures, accessories, and even emotions, which are then incorporated into the generated visualizations. This enables a higher level of personalization and creative control over the characters, ensuring that their design is in line with the vision of the director and the narrative needs of the storyboard.

\subsection{{AI-mediated creative collaboration}}
{Outside the domain of filmmaking, several systems have shown how generative AI can act as a boundary object that helps multiple creative stakeholders externalise ideas and converge on a shared vision. Artinter supports artists and their clients during art-commission negotiations by expanding mood-board concepts with AI-generated examples \cite{chung2023artinter}. We-toon bridges writers and illustrators in the highly iterative web-toon workflow, translating textual feedback into concrete sketch revisions \cite{ko2022we}. RoomDreaming enables homeowners and interior designers to rapidly iterate over photorealistic room variations during early-stage brainstorming \cite{wang2024roomdreaming}. CineVision extends this line of work to the director–cinematographer dyad, integrating shot composition, lighting, style, and character controls in a single real-time interface.}

\section{CINEVISION: DESIGN PROCESS \& GOALS}

In the filmmaking process, visualizing textual information and effectively communicating shot instructions and visual requirements to key departments such as cinematography, costume design, makeup, and art direction is crucial. Traditional communication often relies on the director's ability to convey these ideas through sketches, which can lead to misunderstandings and repetitive clarifications. For example, aspects such as shot angles, lighting design, overall style, and character designs often require extensive back-and-forth between the director and the cinematographer. This communication process is time-consuming and costly. 
 
To address the aforementioned challenges, we conducted user-centered iterative design with directors and cinematographers, resulting in the creation of \textit{CineVision}, a system that supports directors and cinematographers in storyboard creation during the pre-visualization stage. With \textit{CineVision}, our goal is to establish a more efficient and clear visual communication method between directors and cinematographers. The design process comprises three phases: 1) Identifying user pain points and needs: conducting interview studies with experienced directors and cinematographers to uncover strategies they use when communicating visual concepts through storyboards. (Section \ref{Experts Interview}). 2) Prototyping and walkthroughs: designing and developing \textit{CineVision} based on design goals and user requirements derived from the interviews, then obtaining user feedback and iterating accordingly (Section \ref{CINEVISION System Implementation}). 3) Deployment and Evaluation: Conducting user studies to assess how directors and cinematographers interact with \textit{CineVision}, as well as their perceived usability and ease of use (Section \ref{method}).

In this section, we present the first phase of the design process and report the strategies and design goals guiding the design and development of \textit{CineVision} (Table \ref{table1}).

\begin{table*}[ht]
\caption{Film storyboard pre-visualization generation system design summary: of the challenges and strategies reported in Section \ref{Experts Interview}, the design goals derived in Section \ref{design goals}, and the functionalities implemented in CineVision (Section \ref{CINEVISION System Implementation}).}
\scalebox{1}{
\begin{tabular}{llll}
\hline
\multicolumn{1}{c}{Challenges}                                                                        & \multicolumn{1}{c}{Strategies}                                                                                 & \multicolumn{1}{c}{Design Goals}                                                   & \multicolumn{1}{c}{Key Insights}                                                          \\ \hline
\begin{tabular}[c]{@{}l@{}}C1. Complex text-to-shot \\ description\end{tabular}                       & \begin{tabular}[c]{@{}l@{}}S1. Hierarchical input \\ strategy\end{tabular}                                     & \begin{tabular}[c]{@{}l@{}}DO. Lower the creation \\ threshold\end{tabular}       & K1. Tiered menu input                                                                     \\
\begin{tabular}[c]{@{}l@{}}C2. Consistency maintenance \\ across multiple shots/scenes\end{tabular}   & \begin{tabular}[c]{@{}l@{}}S2. Integration with film \& \\ television database matching\end{tabular}           & \begin{tabular}[c]{@{}l@{}}DO2. Ensuring visual \\ continuity\end{tabular}         & \begin{tabular}[c]{@{}l@{}}K2. Text-to-film database \\ matching\end{tabular}             \\
\begin{tabular}[c]{@{}l@{}}C3. Time-consuming repeated \\ confirmations with departments\end{tabular} & \begin{tabular}[c]{@{}l@{}}S3. Providing visual \\ real-time adjustments\end{tabular}                          & \begin{tabular}[c]{@{}l@{}}DO3. Improving communication \\ efficiency\end{tabular} & \begin{tabular}[c]{@{}l@{}}K3. Real-time re-lighting \\ and style adjustment\end{tabular} \\
\begin{tabular}[c]{@{}l@{}}C4. Need for creative exploration \\ and flexible iteration\end{tabular}   & \begin{tabular}[c]{@{}l@{}}S4. Providing flexible \\ customization for \\ characters and costumes\end{tabular} & \begin{tabular}[c]{@{}l@{}}DO4. Supporting creative \\ exploration\end{tabular}    & \begin{tabular}[c]{@{}l@{}}K4. Character design and \\ costume generation\end{tabular}    \\ \hline
\end{tabular}
}
\Description{This table summarizes the relationship between challenges, strategies, design goals, and key insights in the context of a creative production or pre-visualization process.

Challenges (C): These are the core difficulties faced in the process, including complex text-to-shot descriptions, maintaining consistency across shots, the time-consuming nature of repeated confirmations, and the need for flexible iteration and creative exploration.

Strategies (S): These strategies address the challenges. For instance, hierarchical input strategy (S1) helps simplify the input process, and integration with a film & television database (S2) ensures consistency and eases shot descriptions. Real-time visual adjustments (S3) and customization of characters and costumes (S4) foster creative flexibility.

Design Goals (DG): These goals aim to resolve the challenges and align with the strategies. For example, lowering the creative threshold (DO1) ensures that the user can focus on creativity rather than technical complexities. Ensuring visual continuity (DO2) and improving communication efficiency (DO3) are important for smooth collaboration.

Key Insights (K): These insights reflect the lessons learned or key features from the system design that could enhance the process, such as a tiered menu input (K1), text-to-film database matching (K2), real-time re-lighting and style adjustment (K3), and character and costume design generation (K4).

This table offers a holistic view of how challenges are addressed with targeted strategies to achieve specific design goals, ultimately driving key insights that enhance the creative process in film production or similar projects.}
\label{table1}
\end{table*}

\subsection{Experts Interview} 
During the development of \textit{CineVision}, we interviewed two senior industry professionals—one director and one cinematographer—each with over ten years of experience, to deeply understand the practical needs and pain points in the pre-visualization stage. Expert 1, a  director, has extensive experience directing various films and TV shows, while Expert 2, a cinematographer, has been involved in numerous film productions. Throughout our research, we conducted detailed interviews with both experts to assess the challenges they face when creating storyboards during pre-visualization and to understand their strategies. All interviews were recorded and transcribed, and we utilized inductive and deductive methods to analyze the data.

\subsection{Experts Interview Results} \label{Experts Interview}
Both experts emphasized several key attributes necessary for effectively conveying visual language, particularly in the storyboard and shot-planning phases. These attributes include the scene setup, lighting effects and color, character styling, environmental atmosphere, actor dialogues, and the overall emotional tone conveyed by the visuals.

\subsubsection{Hierarchical Input (Key Insights 1, K1)}
Based on interviews with two senior experts, we have identified a key issue: verbal or textual descriptions often result in ambiguity and miscommunication due to their abstract nature (Challenges 1, C1). The interviewed director highlighted difficulties in precisely conveying visual details like shot composition and lighting effects through text, leading to increased communication costs and frequent misunderstandings across departments. To address this, we propose a hierarchical input strategy. This strategy breaks down complex visual descriptions into two distinct layers: a fundamental layer (covering basic visual style elements such as scene type, time of day and lighting direction) and a detail and style layer  (involving clothing, character details, etc.) (Strategies 1, S1). The cinematographer recommended employing standardized terminology and preset templates within the system to minimize ambiguity. Both experts emphasized the value of real-time visual previews, which allow for quick adjustments and refinements of elements, thereby reducing the need for extensive revisions and clarifications.

\subsubsection{Film Database Matching (K2)}
The interviews revealed that maintaining consistency in key visual elements—such as character appearances, lighting, and backgrounds—across multiple shots or scenes was particularly challenging, often disrupting overall visual coherence (C2). The director emphasized that inconsistencies in character styles (design and character) and environment style (lighting and color), especially in complex narratives or multi-character dialogues, required significant additional communications and adjustments. To mitigate this issue, we adopted the strategy proposed by Rao et al. of integrating film database matching \cite{rao2024scriptviz}. This strategy involves automatically retrieving filmic references closely matching the script descriptions and using AI-driven regeneration techniques to ensure visual consistency across scenes~(S2).

\subsubsection{Visual Real-time Adjustment (K3)}
Both the director and cinematographer reported that traditional hand-drawn sketches and verbal descriptions frequently caused ineffective communication, prolonged confirmation processes, and misunderstandings among production departments, slowing down pre-production significantly (C3). The director particularly emphasized that iterative communication not only incurred high time costs but also increased the risk of misinterpretation. In response, we introduced a visual real-time adjustment strategy, enabling directors to directly adjust backgrounds, lighting, and character designs within the system and instantly preview the results. This approach allows immediate feedback and rapid adjustments, streamlining the pre-production workflow (S3).

\subsubsection{Flexible Customization of Roles and Costumes (K4)}
The director and cinematographer both indicated a strong need for frequent experimentation with various environment styles (lighting and color) and shot languages to achieve optimal creative expression. Character costumes and design play a crucial role in conveying the director's vision and tone, as they visually communicate key aspects of a character’s personality, background, and emotional state, facilitating a shared understanding between the director and cinematographer in achieving the desired cinematic language. However, traditional tools typically offer limited flexibility, hindering rapid creative iteration and restricting visual diversity (C4). To address this, the director suggested incorporating more flexible customization options to freely adjust character facial features, costume materials, and colors. Consequently, we developed a flexible customization strategy for roles and costumes, leveraging diffusion models to generatively create and customize diverse materials, enabling dynamic replacements and targeted modifications (S4).

\subsection{{Design Objectives}}\label{design goals}

{To operationalise the four strategies (S1–S4) that emerged from our formative study we distilled four Design Objectives (DO), as shown in Table \ref{X}. Each objective is numbered (O1–O4), explicitly linked to its rationale, and flagged as being rooted in either (U) user‑needs analysis or (P) prior research.}

\begin{table*}[ht]
\caption{{Design Objectives derived from our formative study (U) and prior research (P).}}
\scalebox{0.95}{
\begin{tabular}{cllc}
\hline
  & \multicolumn{1}{c}{ Objective}                                                                                                                                                                             & \multicolumn{1}{c}{ Primary Rationale}                                                                                                                                                                                                              &  Source \\ \hline
 DO1 &  \begin{tabular}[c]{@{}l@{}}Lower the creative threshold. \\ CineVision should let novices externalise ideas\\ without writing long, precise prompts.\end{tabular}                                         &  \begin{tabular}[c]{@{}l@{}}Hierarchical Input strategy reduces ambiguity by surfacing scene, \\ lighting and character menus that mirror how \\ experts iterate in Photoshop/AI editors.\end{tabular}                                              &  U     \\
 DO2 &  \begin{tabular}[c]{@{}l@{}}Ensure visual continuity.\\ Generated shots must maintain background, \\ lighting and character identity across a scene.\end{tabular}                                         &  Visual inconsistency is noted in storyboard literature \cite{rao2024scriptviz}.                                                                                                                                                  &  P      \\
 DO3 &  \begin{tabular}[c]{@{}l@{}}Improve collaboration efficiency. \\ The system should reduce re‑explanation loops \\ between director and cinematographer.\end{tabular}                                       &  \begin{tabular}[c]{@{}l@{}}Real‑time previews and a shared control vocabulary (menus + sliders) \\ let dyads converge on shot angles ``in seconds''.\end{tabular}                                                                                  &  U      \\
 DO4 &  \begin{tabular}[c]{@{}l@{}}Support creative exploration.\\ Provide a curated catalogue of common dialogue \\ compositions and director‑style \\ presets so users can audition looks quickly.\end{tabular} &  \begin{tabular}[c]{@{}l@{}}Lower‑effort reuse of proven framings replaces manual \\ Google‑Image scavenging highlighted by participants;\\ aligns with pattern-based shot design guidelines \cite{glebas2012directing}.\end{tabular} &  P      \\ \hline
\end{tabular}
\Description{Table 2 presents four numbered Design Objectives (DO1–DO4), each with its primary rationale and evidence source. DO1 aims to lower the creative threshold: CineVision should let novices externalize ideas without lengthy prompts, because a hierarchical input menu mirrors expert Photoshop / AI-editor workflows (source U: user study). DO2 focuses on ensuring visual continuity: generated shots must keep consistent background, lighting, and character identity across a scene, addressing visual-inconsistency issues noted in storyboard literature (source P: prior work). DO3 seeks to improve collaboration efficiency: real-time previews plus a shared control vocabulary reduce re-explanation loops between director and cinematographer, letting dyads agree on shot angles “in seconds” (source U). DO4 supports creative exploration: a curated catalogue of common dialogue compositions and director-style presets lowers effort by replacing manual image searching and aligns with pattern-based shot-design research (source P).}
}
\label{X}
\end{table*}

\subsubsection{Baseline workflows}

{For comparison, we define (i) a manual sketch‑plus‑photo‑editing pipeline that predates generative AI and (ii) a single‑prompt diffusion workflow (e.g., Midjourney) in which users type one long prompt and often omit critical details. Both baselines suffer from the visual inconsistency described above.}

\subsubsection{Lower the Creative Threshold {(DO1)}}
Based on our interviews, directors and cinematographers often face C1, the challenge of translating abstract textual descriptions into precise shot compositions. They reported that relying solely on natural language leads to ambiguous and incomplete visual representations. To address this, our strategy (S1) employs a hierarchical input system that breaks down the overall vision into distinct elements—such as scene type, basic lighting, and preliminary character positions—using standardized menus and pre-set options. By leveraging the comprehensive film database, the system automatically matches script elements with appropriate visual references. Furthermore, integrating automated character design and costume generation via diffusion models significantly reduces the burden of manual detailing. Collectively, this approach fulfills {DO1}: lowering the creation threshold by simplifying the process, ensuring that directors can easily externalize their ideas into coherent storyboard visuals without getting stuck in technical intricacies.
{While the menu system constrains wording diversity, participants valued its speed over expressive breadth, especially novices unfamiliar with cinematography jargon.}

\subsubsection{Ensure visual continuity {(DO2)}}
One major challenge identified in our research is C2: maintaining visual consistency across multiple shots and scenes. Directors noted that traditional storyboards often lack continuity, resulting in discrepancies in character appearance, lighting, and scene composition. {For example, prompting ``a man in a denim jacket at dusk'' three times in a single‑prompt diffuser produced three different jacket hues and skyline colours, illustrating how even identical prompts can yield disruptive visual drift.} To overcome this, our strategy (S2) incorporates automated semantic matching using a comprehensive film database. This process matches shots based on time sequences and character dialogues, emulating the natural progression of a scene from framing and shot composition to capturing the dynamics of dialogue and interaction \cite{rao2024scriptviz}. Additionally, our system applies advanced AI-driven relighting techniques to ensure that lighting, its direction, type, texture, and color, remain consistent throughout the storyboard. Together, these approaches achieve {DO2}, ensuring visual continuity, by providing a seamless and coherent visual narrative that faithfully reflects the director's intent on all shots.

\subsubsection{Improve communication efficiency {(DO3)}}
A recurring pain point identified in our interviews was C3, referring to the inefficiency caused by repeated confirmations and clarifications among directors, cinematographers, and other production departments. To address this, our strategy (S3) emphasizes real-time visual adjustments and immediate feedback. With real-time preview capabilities, directors can instantly adjust lighting and stylistic parameters and see their effects directly reflected on the storyboard. Additionally, the system allows users to explore diverse director-specific styles via the ``Director Master Style'' feature and integrates inputs from a film database. This ensures that all team members have clear and shared references, directly advancing {DO3} by reducing unnecessary iterative interactions, enhancing communication efficiency, and enabling stakeholders to swiftly refine visual details and progress smoothly through the production process.

\subsubsection{Support creative exploration {(DO4)}}
Finally, our research highlighted C4: the need for flexible, iterative creative exploration. Directors and cinematographers expressed the desire to rapidly experiment with different character designs and emotional atmospheres, enabling the exploration of diverse visual combinations without being constrained by rigid workflows. In response, our strategy (S4) provides extensive customization options. Leveraging diffusion models and insights from industry experts, the system adjusts costumes and incorporates features such as actor count control, precise spatial positioning, and mood-specific adjustments. Together, these features realize {DO4} (supporting creative exploration), enabling filmmakers to rapidly iterate and experiment with various aesthetic possibilities, thus achieving an optimal balance between creative freedom and production feasibility.

\section{CINEVISION SYSTEM IMPLEMENTATION} \label{CINEVISION System Implementation}
The \textit{CineVision} system is designed to provide an efficient and coherent pre-visualization process, transitioning seamlessly from script to storyboard. The system integrates three core modules: (1) Relighting and Stylization, (2) Character Design and Costume Creation, and (3) Text-to-Film Database Matching. These modules are brought together in a simplified, two-tier menu system that guarantees the final storyboard not only aligns with the director’s creative vision but also maintains high visual consistency and artistic quality.

\begin{figure*}[ht]
    \centering
    \includegraphics[width=18cm]{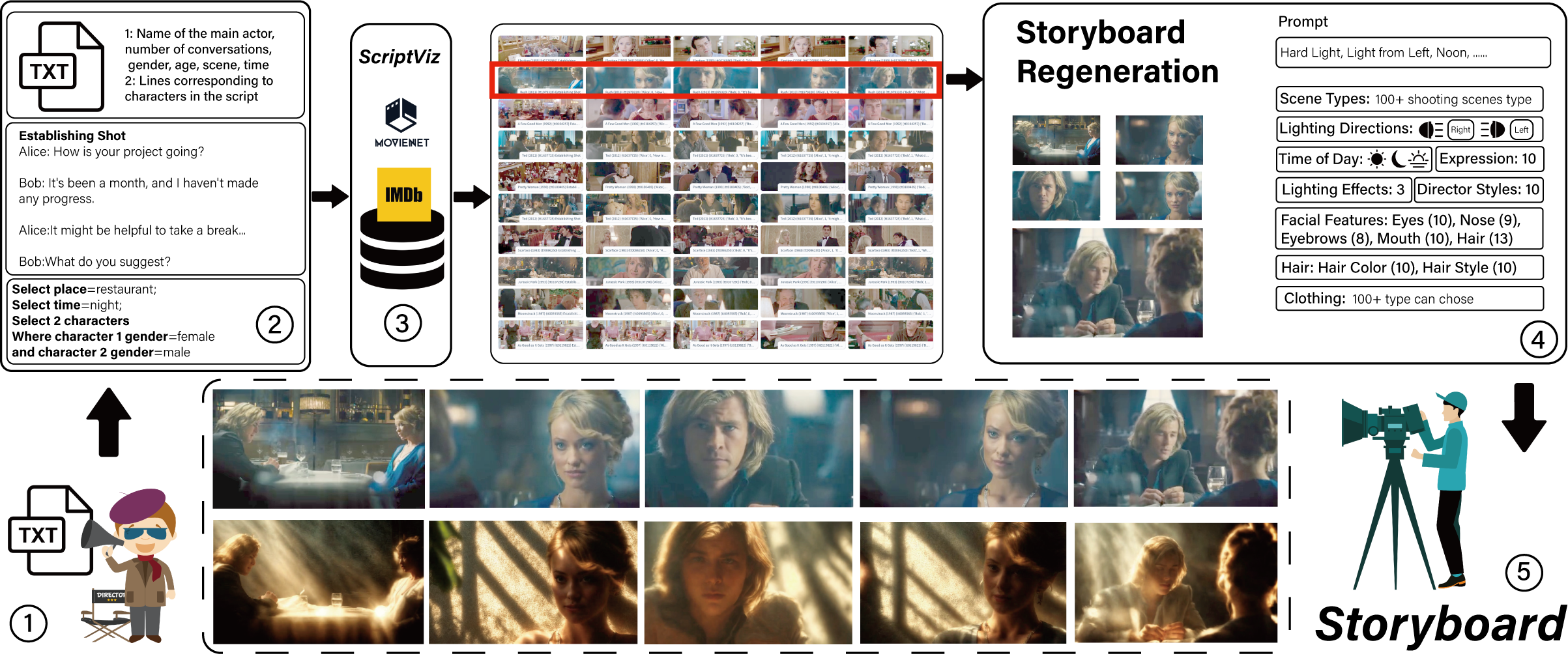}
    \caption{The \textit{CineVision} System consists of five fundamental components: (1) \textbf{Text Information Input}: The director inputs the script information into the system. (2) \textbf{Script Information Filtering}: Based on the director's vision, the director provides specific details for the dialogue scenes, such as the number of characters, their genders, time of day, and the location where the dialogue occurs. (3) \textbf{Text to Film Database Match}: The system searches the film database for storyboards that match the given conditions. (4) \textbf{Film Storyboard Regeneration and Stylization}: The director defines the visual parameters, including scene composition, light direction, lighting effects, facial expressions of the actors, and the actors' costumes and makeup. In addition, the director can also choose the environment style (lighting, color) of a well-known director to generate a customized storyboard, thereby achieving stylized customization. (5) \textbf{Cinematographer Execution Phase}: Approved storyboards are sent to the cinematographer, conveying the director's vision and guiding lighting and camera setups.}
    \Description{The \textit{CineVision} System consists of five fundamental components: (1) \textbf{Text Information Input}: The director inputs the script information into the system. (2) \textbf{Script Information Filtering}: Based on the director's vision, the director provides specific details for the dialogue scenes, such as the number of characters, their genders, time of day, and the location where the dialogue occurs. (3) \textbf{Text to Film Database Match}: The system searches the film database for storyboards that match the given conditions. (4) \textbf{Film Storyboard Regeneration and Stylization}: The director defines the visual parameters, including scene composition, light direction, lighting effects, facial expressions of the actors, and the actors' costumes and makeup. In addition, the director can also choose the environment style (lighting, color) of a well-known director to generate a customized storyboard, thereby achieving stylized customization. (5) \textbf{Cinematographer Execution Phase}: Approved storyboards are sent to the cinematographer, conveying the director's vision and guiding lighting and camera setups.}
    \label{pipeline}
\end{figure*}

\subsection{Core System Components}

\subsubsection{Relighting and Stylization}
Lighting plays a crucial role in establishing the emotional tone and atmosphere of a scene, as described in the script \cite{graham2022play, matbouly2022quantifying}. 
Inspired by IC-Light~\cite{zhang2024ic}, we developed our relighting and stylization capabilities by finetuning a pretrained Stable Diffusion model 1.5 (SD1.5)\footnote{\url{https://huggingface.co/stable-diffusion-v1-5/stable-diffusion-v1-5}} {and then empowering it with relighting capabilities. The finetuning is conducted on a dataset of 1,000 movie images from \textit{MovieNet} \cite{huang2020movienet} to adapt the model to the movie domain.} The dataset consists of scenes from Hollywood films, focusing on dialogue-driven moments between two characters in well-lit settings. The images feature dynamic lighting setups such as high-key, backlighting, and naturalistic lighting, ensuring a range of styles for the model to learn. We prioritize scenes with clear character interactions, balanced compositions, and high visual appeal, including intimate dialogues and emotionally charged exchanges, which allow for nuanced relighting and style adaptations.
{Two researchers manually screened \textit{MovieNet}’s still frames and independently tagged those that met four criteria: (1) exactly two speaking characters in close-up or medium shots; (2) no blur or exposure issues; (3) diverse lighting setups—from high-key and low-key to back-lighting and natural light; and (4) balanced composition that preserves spatial context. The two candidate lists (approximately 1,200 frames) were reconciled in a joint review, producing a curated set of 1,000 images for further fine-tuning.}

We use \textit{ChatGPT4o} to describe the lighting and style of these images to form text-image pairs. 
With these 1,000 pairs, we finetune the SD1.5 for 5 epochs on a single NVIDIA A100 80GB GPU. Then we follow the IC-Light~\cite{zhang2024ic} to take the above finetuned SD1.5 as our foundation model and empower it with relighting ability.
{This model optimizes intra-scene lighting consistency, effectively managing multiple light sources and ensuring smooth transitions in user-selected scene with multiple characters. }

Directors can emulate various filmmaking styles, such as those of Martin Scorsese or Stanley Kubrick, by applying pre-configured prompts that unify lighting, color tones, and character design. To develop these director-specific style prompts, one of our authors—who has five years of film production experience—selected 10 renowned filmmakers, analyzed a representative work from each, and extracted five key shots to form a reference library. Guided by instructions like, ``This is a shot from Wes Anderson's film. Help me generate a description of the shot in their style, focusing on lighting, color, and character style (design and character),'' {the \textit{ChatGPT4o} Large Vision Model \cite{zhu2023minigpt} was repeatedly applied.} After three rounds of testing and optimization, we finalized unique style descriptions for each director. This approach ensures that visual outputs remain consistent with the director’s cinematic vision, enhancing communication efficiency (DO3 : Improve communication efficiency).

\subsubsection{Character Design and Costume Creation}
Character design and costume creation are essential to storytelling, yet traditional storyboard methods often treat them separately from shot planning, leading to inconsistencies or delays in visualizing a character’s on-screen appearance. \textit{CineVision} addresses this by integrating character design and costume creation directly into the pre-visualization process by further finetuning the above SD1.5. The system generates highly customizable character traits and costume designs based on paired text-to-image data, including detailed character descriptions and costume features, ensuring that the designs align with the scene's overall visual direction and tone. 
We further fine-tuned the above SD1.5 to enhance the accuracy and customization of character designs using a dataset of 2,000 images from \textit{MovieNet}. 
{Two researchers independently browsed the corpus and selected close-up or medium portraits that (i) clearly show the face and upper body, (ii) cover a wide range of costume eras—from period pieces to contemporary and fantasy styles, (iii) reflect varied body types, ages, and emotional states, and (iv) exhibit good exposure and sharpness. The final set, reached by consensus, provides balanced coverage of leading, supporting, and background roles across diverse narrative contexts.} These images primarily feature close-up and medium-shot portraits of characters in diverse costume styles, ranging from period pieces to modern attire and fantasy elements. We prioritized clear facial expressions, varied body types, and distinctive design elements, ensuring coverage of different character roles, emotional states, and narrative contexts.

We use \textit{ChatGPT4o} to describe the character and costume of these images to form text-image pairs. With these pairs, we finetune the SD1.5 for 5 epochs on a single NVIDIA A100 80GB GPU. It enables directors to adjust facial features, hairstyles, and costume details such as type, texture, and color, ensuring harmony with shot composition and lighting. 
{Similar to the above relighting and stylization, we fine-tuned SD 1.5 weights to achieve better character and costume performance.}

The model is able to handle localized lighting adjustments for characters and costumes, ensuring continuity in lighting direction and intensity across shots, even with changing character poses and interactions (DO4 : Support creative exploration ).

\subsubsection{Text-to-Film Database Matching}

In traditional storyboard creation processes, scenes are typically divided into multiple shots, including dialogues, varied angles, and detailed compositions. However, existing systems, such as those by Shi et al. \cite{shi2020emog} and Rusu et al.\cite{rusu2024script}, often lack visual continuity across shots, providing outputs suitable only as rough references and failing to capture the detailed shot compositions required for complete scenes. Building upon \textit{Scriptviz}~\cite{rao2024scriptviz}, \textit{CineVision} optimizes database matching algorithms to better manage the challenges posed by multi-character interactions. We similarly used Structured Query Language (SQL) \cite{jamison2003structured}, enabling users to specify attributes clearly defined in their vision and highlight differences in uncertain attributes. 

{To further support cross-shot coherence, we introduce a scene-complexity filter that automatically skips shots containing more than six visible characters (typical of crowd or group-opening scenes). Such ensembles make subsequent lighting and style transfers unreliable, increasing the risk of visual discontinuities. By deprioritising these high-complexity shots, the system delivers more consistent lighting, stylisation, and character appearance across adjacent frames—an incremental but measurable improvement toward visual continuity rather than a complete guarantee.}

\subsubsection{Hierarchical Menu Input}
In \textit{CineVision}, the input for generating storyboards is structured hierarchically, reflecting how directors and cinematographers typically communicate scene details. This hierarchical system divides prompts into primary and secondary categories, allowing for more intuitive input and reducing ambiguity. The primary categories (e.g., scene environment and lighting and director's style) are defined in the first-tier menu, while more specific details (e.g., character features and clothing and facial expressions) are included in the second-tier menu. To enhance the accuracy and coherence of generated storyboards, each category in the menu system is assigned a weighted value. These weights are based on the relative importance of each element in contributing to the overall visual composition. Specifically, the weights for the first-tier menu are higher than those for the second-tier menu. Within the first-tier, environmental settings, time of day, the basic attributes of the actors, and director's style are weighted more heavily than other categories. In the second-tier, character facial detail features (including style, facial expressions, type, and color) are weighted more heavily than other categories. The weighted values for each prompt category are then combined to form a final weighted input that influences the text,to image generation process (DO1 : Lower the creation threshold),

\[
W_{\text{total}} = \sum_{i=1}^{n} w_i \cdot x_i
\]

\( W_{\text{total}} \) is the total weighted input for generation, \( w_i \) is the weight of the \( i^{\text{th}} \) category (e.g., environment, lighting), \( x_i \) is the input value for that category, and \( n \) is the total number of categories (first-tier and second-tier).

\subsection{System Workflow}
\textit{CineVision} offers an efficient and simplified method for matching script descriptions with actual storyboard frames and regenerating the creation. As shown in Figure \ref{pipeline}, the system is divided into the following five parts: (1) Text Information Input; (2) Script Information Filtering; (3) Text to Film Database Match; (4) Film Storyboard Regeneration and Stylization; (5) Cinematographer Execution Phase.

\begin{figure}[ht]
    \centering
    \includegraphics[width=8.5cm]{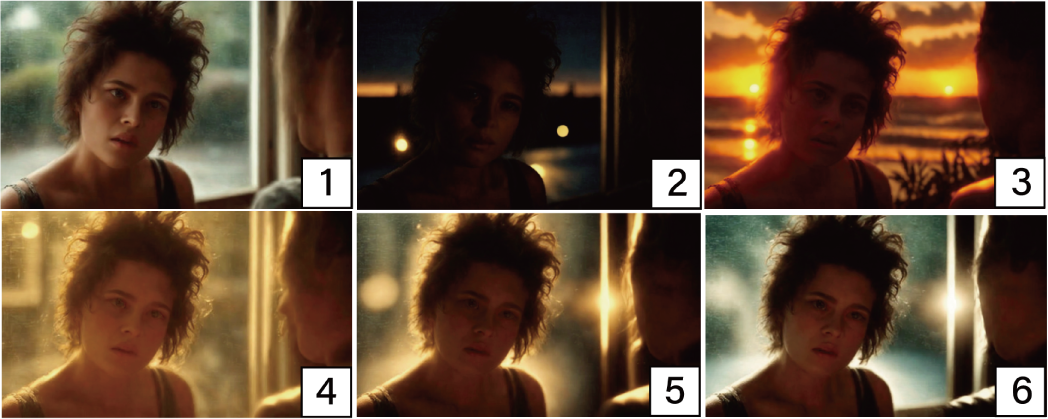}
    \caption{The Relighting: (1) Noon. (2) Night. (3) Sunrise or Sunset. (4) Soft Light (5) Hard Light (6) Key Light.}
    \Description{The Relighting: (1) Noon. (2) Night. (3) Sunrise or Sunset. (4) Soft Light (5) Hard Light (6) Key Light.}
    \label{light}
\end{figure}

\subsubsection{Relighting and Stylization}
In the \textit{CineVision} system, the Relighting and Stylization features are core tools for enhancing cinematic visual effects and accurately conveying the director's intentions. By fine-tuning the lighting and imitating specific directorial styles, directors can precisely set the lighting atmosphere and environment style (lighting, color) of each scene during the pre-production phase, ensuring more efficient communication with cinematographers and contributing to the consistency and artistic quality of the final work.

\textbf{Relighting.} In \textit{CineVision}, the Relighting feature offers reference options that align with cinematic styles. It provides directors with multiple adjustable options to tailor the lighting effects according to script requirements. The system offers three common time-of-day options: ``Noon'' for uniform, intense daylight emphasizing clarity and brightness; ``Night,'' featuring low light sources and cool-toned illumination; and ``Sunrise or Sunset'', which uses warm golden and orange light to create a romantic or emotionally rich atmosphere, as shown in Figure \ref{light} (1)-(3). 

\begin{figure}[ht]
    \centering
    \includegraphics[width=8.5cm]{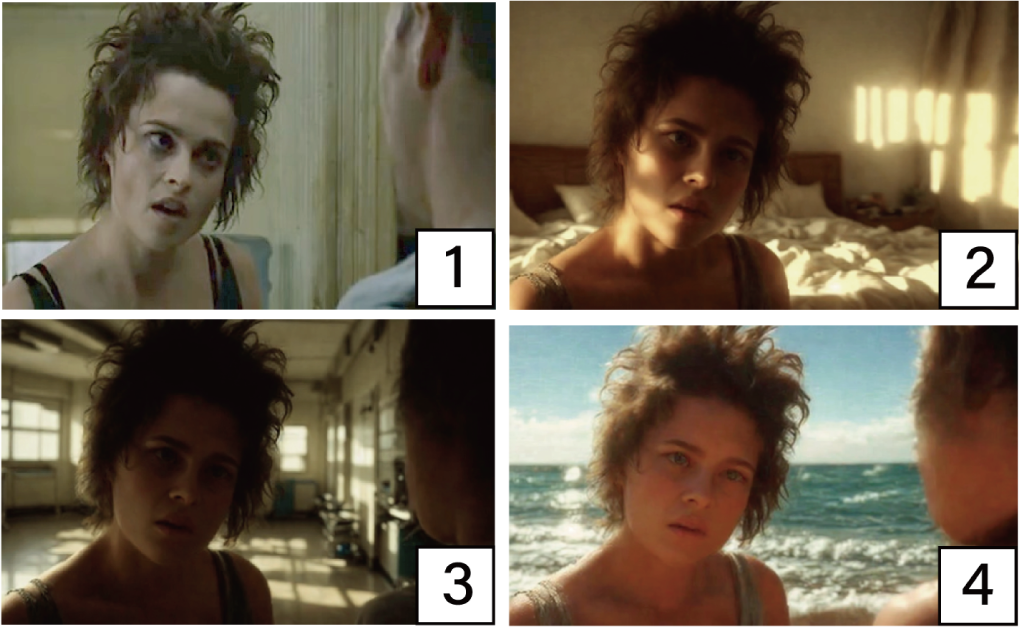}
    \caption{The background scenarios: (1) Original film footage. (2) Bedroom. (3) Indoor corridor. (4) Beaches.}
    \Description{The background scenarios: (1) Original film footage. (2) Bedroom. (3) Indoor corridor. (4) Beaches.}
    \label{Background}
\end{figure}

Regarding light type selection, considering that directors may not be familiar with professional lighting equipment, \textit{CineVision} provides intuitive and easy-to-understand options such as ``Soft Light'', ``Hard Light'', and ``Key Light'', which align with the communication preferences of directors and cinematographers, simplifying the complexity of lighting design, as shown in Figure \ref{light} (4)-(6). Additionally, the system offers simple light source direction choices, such as ``Left Light Source'' or ``Right Light Source'', to help directors establish different lighting effects. To further assist directors in accurately setting the shooting environment, \textit{CineVision} provides over 100 common background scenarios, covering everyday life, indoor, and outdoor environments, ensuring that the lighting effects coordinate with the background to achieve perfect visual outcomes, as shown in Figure \ref{Background} (1)-(4).

\begin{figure}[ht]
    \centering
    \includegraphics[width=8.5cm]{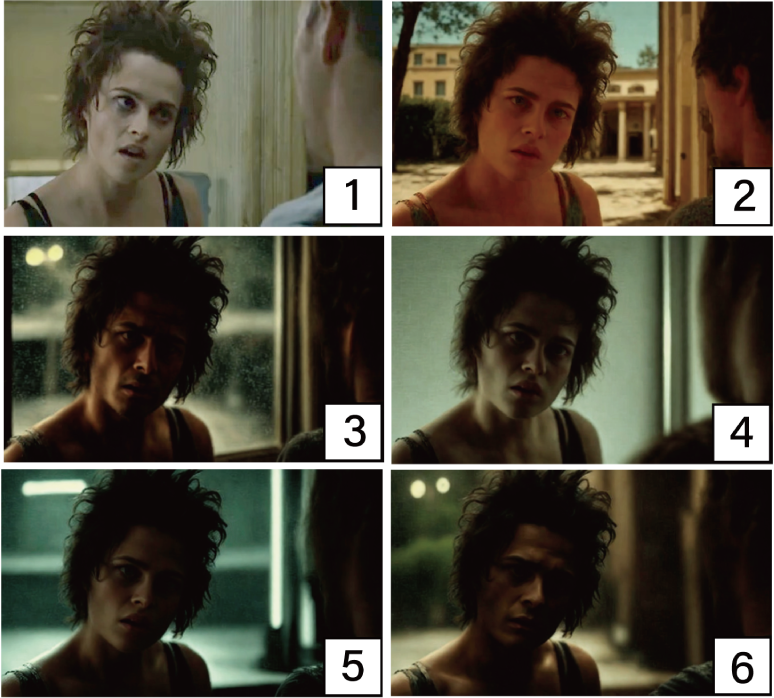}
    \caption{The five director's stylizations: (1) Original film footage. (2) Wes Anderson director style. (3) Martin Scorsese director style. (4) Stanley Kubrick director style. (5) Ridley Scott director style. (6) Russo Brothers director style.}
    \Description{The five director's stylizations: (1) Original film footage. (2) Wes Anderson director style. (3) Martin Scorsese director style. (4) Stanley Kubrick director style. (5) Ridley Scott director style. (6) Russo Brothers director style.}
    \label{director}
\end{figure}

\textbf{Stylization.} Every director has a unique environment style with lighting, color tones, and camera styles often becoming trademarks of their work. To help directors quickly achieve these styles during the creative process, \textit{CineVision} offers a stylization adjustment feature. Through 10 prompt style customizations, directors can choose from various classic directorial styles and apply them to their scenes, thereby precisely shaping the visual atmosphere and enhancing the artistic expression of the film. These finely tuned and verified prompts are presented as buttons with the director’s name and style, allowing users to apply them directly within the system, as shown in Figure \ref{interface} (K). Figure \ref{director} (1)-(6) shows the movie footage and the results of the five directors' stylizations.

\subsubsection{Character Design and Costume Creation}
In the \textit{CineVision} system, Character Design and Costume Creation is a crucial phase for transforming the director’s vision into visually compelling and authentic character portrayals. This process involves selecting the right attributes, leveraging filmmaking expertise, and using diffusion models to create character models and costumes that align with the overall narrative and aesthetic style of the film.

\begin{figure}[ht]
    \centering
    \includegraphics[width=8.5cm]{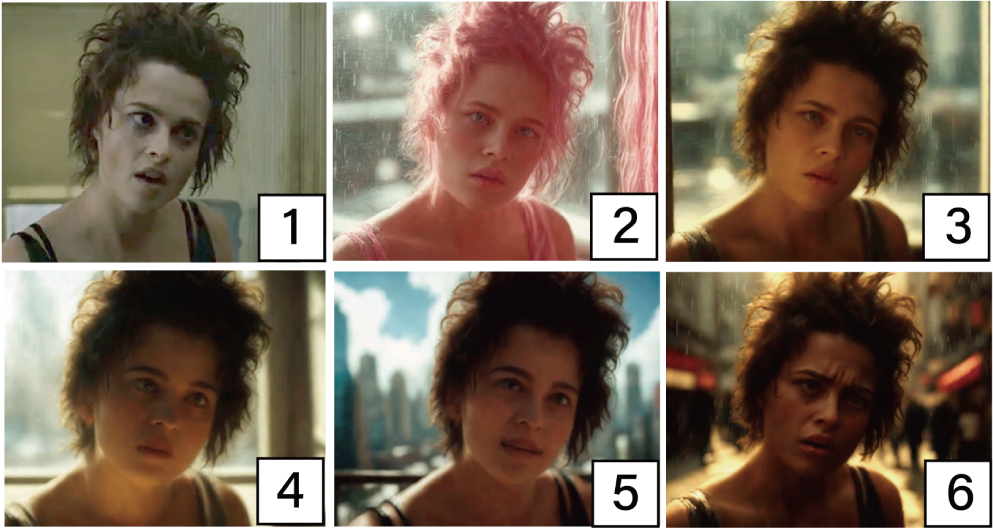}
    \caption{Character Design: (1) Original film footage. (2) Pink Hair \& Tired Eyes. (3) Tired Eyes. (4) Large, Wide Nose. (5) Wide Smile  Mouth. (6) Angry Expression Style.}
    \Description{Character Design: (1) Original film footage. (2) Pink Hair \& Tired Eyes. (3) Tired Eyes. (4) Large, Wide Nose. (5) Wide Smile  Mouth. (6) Angry Expression Style.}
    \label{Character Design}
\end{figure}

\textbf{Character Design.} Creating characters in \textit{CineVision} is a meticulous process that begins with understanding the personality, role, and visual traits of each character in the story. The system allows directors to modify facial features such as expressions, eyes, nose, and mouth based on visual references of the actor, ensuring the character’s appearance aligns with the emotions and personality they are meant to convey. Additionally, the system offers options for adjusting hair characteristics, such as length, style, and color, making it easy to create unique characters that fit the director’s vision and meet audience expectations. Directors can fine-tune individual attributes within the user interface, selecting specific eye shapes, nose contours, and mouth expressions, as shown in Figure \ref{Character Design} (1)-(6). This level of customization enables directors to adjust the appearance of each character to reflect their personality and narrative role, ensuring seamless alignment between the character’s visual design and their function within the story.

\textbf{Costume Design.} Equally important as character design, costume creation further enhances character's identity, role, and the world they inhabit. In \textit{CineVision}, costume design is a flexible and intuitive process. The system provides multiple options for adjusting clothing elements such as tops, pants/skirts, dresses, and other apparel, allowing directors to style characters according to the desired time period, social status, and the thematic tone of the film. Customizable features include fabric textures, clothing colors, and costume styles, all designed to capture the essence of the world and the character’s journey within it. The system also allows directors to customize the overall look by selecting clothing styles (e.g., modern and vintage and formal and casual), colors (e.g., bold, soft, muted), and material textures (e.g., leather and silk and denim), ensuring the character’s appearance reflects both their personality and the scene's atmosphere, as shown in Figure \ref{Character Design} (1)-(4). Once the customization is complete, the system applies the selected outfits to the character in the visual reference, ensuring that the character’s appearance aligns with their personality and the overall mood of the scene.

\begin{figure}[ht]
    \centering
    \includegraphics[width=8.5cm]{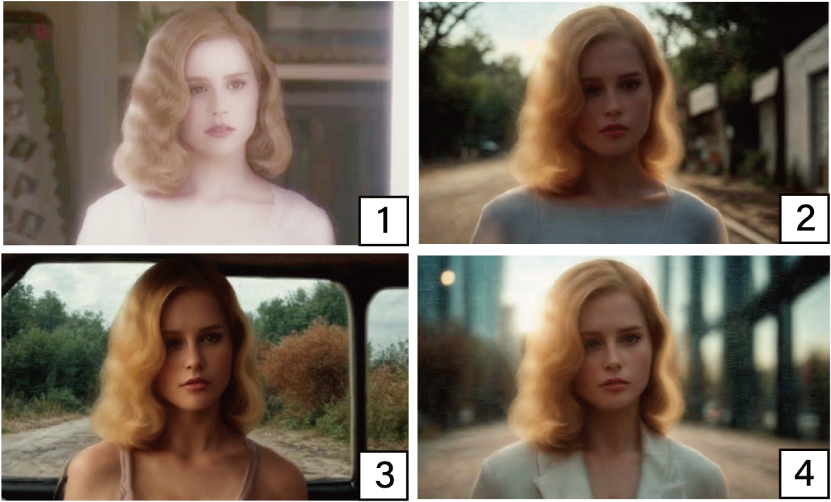}
    \caption{Costume Design: (1) Original film footage. (2) T-shirt. (3) Tank Top. (4) Business Attire.}
    \Description{Costume Design: (1) Original film footage. (2) T-shirt. (3) Tank Top. (4) Business Attire.}
    \label{Character Design}
\end{figure}

\subsubsection{Script-to-Storyboard Conversion}
Traditional storyboard creation often requires directors and cinematographers to provide detailed descriptions for each scene, including actions, positions, camera angles, and other specifics, which can be complex and prone to misunderstandings. Furthermore, in AI-driven storyboard generation systems, using a single prompt for the entire scene can lead to inaccuracies, as the system must process and combine all elements of the prompt to generate an image. This can result in inconsistent visual output, difficulty maintaining character continuity, poor background consistency, and issues with seamless lighting transitions \cite{hu2024animate,xie2023wakey}. To address these challenges, \textit{CineVision} utilizes a simplified two-tier menu system. This system structures prompts based on the different dimensions of communication that directors and cinematographers use in everyday storyboard discussions. By assigning weighted proportions to various prompt descriptions at each menu tier, \textit{CineVision} ensures more accurate and coherent text‑to‑image generation while reducing input complexity. {The preset library currently covers ~180 live‑action dialogue framings, satisfying over 90\% of scenarios surfaced in our formative study; we will continue to expand this catalogue as new use cases emerge.}

\textbf{First-tier menu.} Directors can select the basic attributes of the scene, including environmental settings (e.g., bedroom or airport), light source direction (e.g., left or right), time of day (e.g., noon, night, or sunset), director style (with 10 famous director styles available for selection), and lighting effects (e.g., soft light or hard light, key light). Additionally, the basic attributes of the actors (e.g., expressions such as anger, joy, or sadness, and facial features such as eyes, nose, or mouth) and hairstyle can also be set. To further simplify input, the system provides a list of commonly used and effective descriptive terms for directors to choose from, eliminating the need for detailed manual descriptions.

\textbf{Second-tier menu.} Directors can further refine the details of characters’ facial features, hairstyle length and color, and clothing style (including material and color). This structured and intuitive input method not only simplifies the workflow but also effectively avoids the ambiguity and complexity associated with traditional methods. {Directors can focus more on creative aspects rather than technical details. By selecting the most relevant scene descriptions, the system can help reduce communication overhead and maintain alignment with the director’s intent.}

\subsection{System Setup}

The \textit{CineVision} platform is built and deployed on Ubuntu, accessible via a public URL. It utilizes an \textit{NVIDIA RTX 4090 GPU} with 24GB of memory for inference, and the backend is powered by the \textit{PyTorch} platform. The front-end user interface is based on \textit{Gradio}~\cite{abid2019gradio}, offering an intuitive, user-friendly experience for interacting with models and applications. Our graphical user interface includes a customizable, hierarchical menu system, drag-and-drop functionality, and live feedback. 
It also features interactive galleries, prompt building with contextual suggestions, and advanced image processing options to enhance user interaction. For the generated movie images, the system is set to output at a fixed resolution of $960 \times 536$ pixels.

\textbf{{Implementation Details}}
{
For the scene matching component, we formulate Structured Query Language style queries based on user-defined fixed and variable visual attributes (e.g., location, time of day, gender). These queries are used to filter and retrieve candidate scenes from a preprocessed \textit{MovieNet} database. Each candidate is then scored using CLIP-based visual-text similarity metrics applied to setting tags, along with face recognizability measures. This scoring helps identify the most relevant and visually clear establishing shots and dialogue-aligned frames. For the scene generation component, we follow standard practices from the SD 1.5 model, leveraging a Variational Autoencoder to process images and CLIP-based text encoders to process descriptive prompts. These generation processes leverage the SD 1.5 model fine-tuned as described in Sections 4.1.1 and 4.1.2. And these prompts—detailing background, lighting effects, directorial style, character appearance, and costume design—serve as conditioning inputs for the generation process.}

\textbf{{Tool scope}}. {\textit{CineVision} currently supports only conventional live-action dialogue scenes. The model cannot modify character movements, apply skeletal or facial rig animations, render non-photoreal styles (such as cartoons or sketches), or handle complex camera path control. These limitations arise from training-data bias and hardware constraints.}

\begin{figure*}[ht]
    \centering
    \includegraphics[width=17cm]{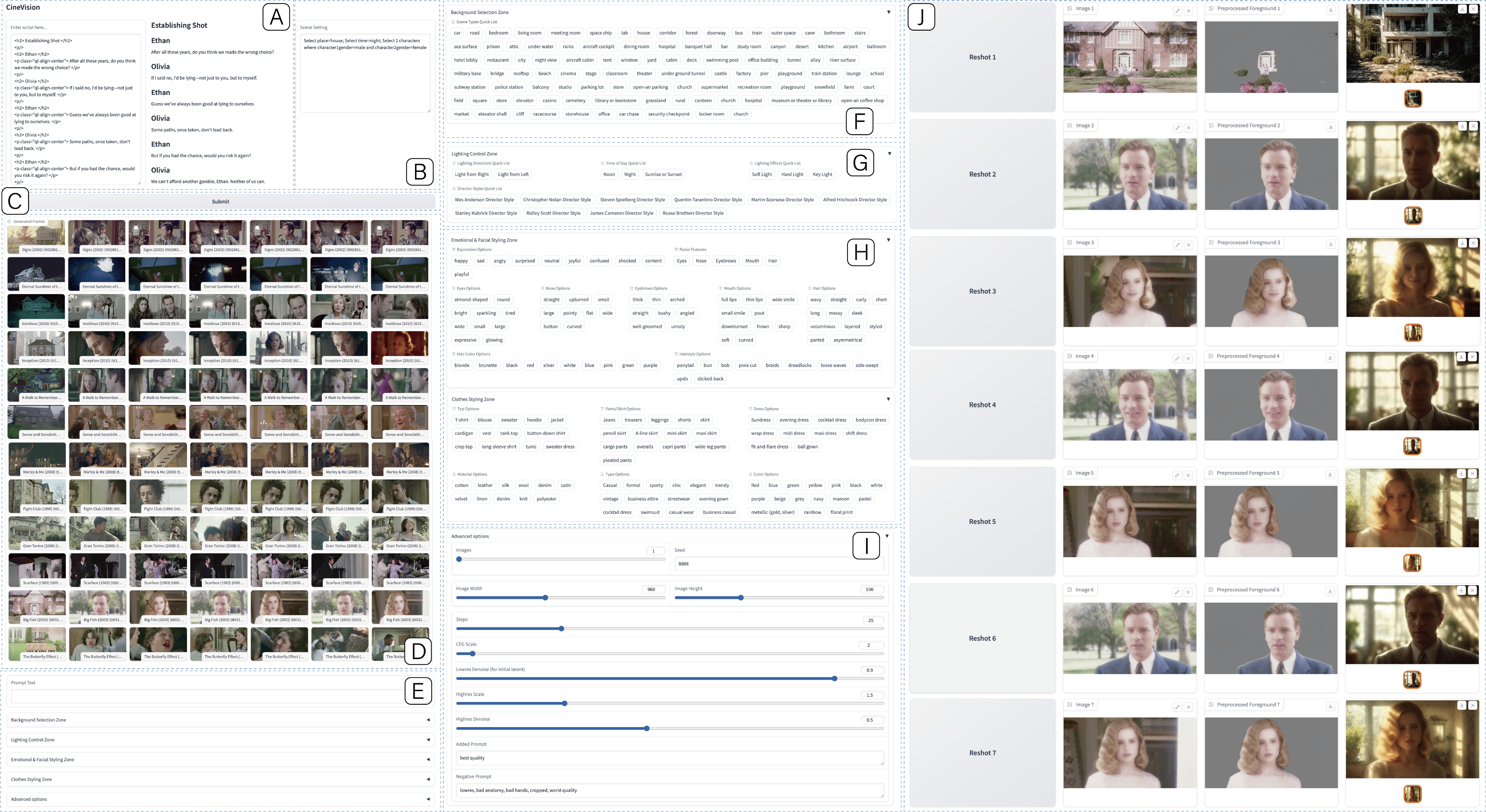}
    \caption{{\textit{CineVision} interface (composite view). Users iteratively refine a storyboard in four stages: (A) edit or revise the script (and optional SQL-like metadata in B); pressing \textit{Submit} (C) retrieves a gallery of matched reference-shot groups (D). The user selects a preferred shot group, pins it to the left column of the real-time preview board (J), and then adjusts prompts (E) as well as background, lighting, and character controls (F–H) plus system parameters (I). Clicking \textit{Reshot} in J regenerates only the pinned frames with the latest settings, while the original remains for side-by-side comparison. At any time the user can return to the script in A/B, resubmit, and restart the loop, enabling non-destructive, flexible iteration from script to image.}}
    \Description{The \textit{CineVision} System Interface: After the user completes the script and attribute configuration in the script editing area (A) and SQL statement area (B), the corresponding image (D) is generated by pressing the submit button (C). Users can enter their own prompt (E). The background, lighting, characters and other elements (F, G, H) can be adjusted in the right-hand area according to different needs. The user can adjust the parameters of the system (I). Finally, the effect can be viewed in the real-time display interface (J), achieving visual and rapid generation and flexible customization from script to image. }
    \label{interface}
\end{figure*}

\subsection{System Interface}
As shown in Figure \ref{interface}, \textit{CineVision} provides a user-friendly interface enabling directors and cinematographers to modify the script in real time, review visuals, and refine individual frames, including adjustments to background, lighting, and character design. This ensures alignment with the dialogue and visual consistency across the project, with all images drawn from the same scene to preserve coherence.

\textbf{Input and Output}: Users input the script (A) along with SQL statements that define both fixed and variable attributes (B). After submission (C), the system generates multiple outputs, each consisting of images that correspond to the dialogue lines within the script (D). Users can enter their own prompt (E) and they can also make selections for background (F), lighting effects \& director styles (G), character design (H), adjust the parameters of the system (I) and real-time display interface (J), as illustrated in Figure \ref{interface}.

\textbf{Modification}: Users can modify images by selecting backgrounds from a library of over 100 options, adjusting lighting effects (e.g., soft light, hard light and key light), and choosing light direction. The system ensures the lighting matches the background for visual consistency. For Character Design and Costume Creation, users can adjust facial features, hairstyles and hair color, and clothing styles (e.g., tops, pants/skirts, classicism). These adjustments allow for the creation of characters that align with the director's vision and the script's requirements.

\section{EVALUATION} \label{method}

\subsection{Dependent Variables}
The primary focus of our measurements was participants' user experience and workload. Upon completion of each experimental condition, participants were asked to complete two questionnaires, as outlined in the following sections. In addition, a 10-minute semi-structured interview was conducted with each participant to collect qualitative feedback.

\subsubsection{NASA-TLX Questionnaire}
To assess participants' workload, we used the NASA-TLX Questionnaire by Hart and Staveland \cite{hart2006nasa}, known for its validation and broad use in measuring subjective workload. The questionnaire includes six dimensions—Mental Demand, Physical Demand, Temporal Demand, Performance, Effort, and Frustration—each rated on a 7-point Likert scale. These scores are combined to provide an overall workload score, offering insight into the cognitive and physical demands of the task. A score of 7 indicates the highest perceived workload, while a score of 1 reflects the lowest workload.

\subsubsection{User Experience Questionnaire}
We used the User Experience Questionnaire (UEQ) to assess participants' overall experience, focusing on the system's usability and effectiveness in storyboard creation and facilitating communication between the director and cinematographer. The questionnaire uses a 7-point Likert scale, where a score of 7 indicates the highest experience and a score of 1 reflects the lowest experience. It covers three areas: ``Overall Experience'', ``Usefulness'', and ``Ease of Use''. Adapted from prior studies \cite{rao2024scriptviz,xu2024transforming}, it evaluates participants' adaptation to the storyboard creation process, complementing the NASA-TLX.

\subsubsection{Observation and Semi-structured interviews}
During the experiment, we observed participants' interactions with \textit{CineVision}. Afterward, all participants underwent semi-structured interviews to provide detailed feedback on their experience using the system. These qualitative insights complemented the quantitative data. Participants were asked key questions regarding their production outcomes and experiences, including: (1) How did the system impact your efficiency in pre-visualization and storyboard creation? (2) What challenges or limitations did you encounter when using \textit{CineVision} for scene composition and lighting adjustments? (3) How did the ability to customize environment style (lighting, color) and character style (design and character) details affect your creative process?

\subsection{Participants}
We recruited 24 participants through social media, who provided self-reported demographic information, including gender (12 females and 12 males) and age ($ \bar{M}$ = 25.42, SD = 1.99). All participants reported normal or corrected vision. {Six participants had more than one year of experience in the film industry (3 directors and 3 cinematographers).} The remaining 18 participants were amateurs who had tried their hand at creating short films or shooting short videos. As shown in Table \ref{group}, the participants were divided into three groups: Group A (using \textit{CineVision}), Group B (using AI tools for creation \textit{DALL·E 3}), and Group C (using traditional hand-drawn storyboards) as the baseline condition.

\begin{table}[ht]
\center
\caption{Depending on the level of experience and the tools used, participants were assigned to one of three groups: Group A (using \textit{CineVision}), Group B (using AI tools), and Group C (using traditional hand-drawn storyboards). In each group, D1-D3 were professional directors, AD1-AD9 were amateurs (directors), C1-C3 were professional cinematographers, and AC1-AC9 were amateurs (cinematographers).}
\scalebox{0.9}{
\begin{tabular}{ccccc}
\hline
        & Team 1   & Team 2   & Team 3     & Team 4     \\ \hline
Group A & (D1\&C1) & (AD1\&AC1) & (AD2\&AC2) & (AD3\&AC3) \\
Group B & (D2\&C2) & (AD4\&AC4) & (AD5\&AC5) & (AD6\&AC6) \\
Group C & (D3\&C3) & (AD7\&AC7) & (AD8\&AC8) & (AD9\&AC9) \\ \hline
\end{tabular}
\label{group}
\Description{This table outlines the structure for participant assignment based on the tools used and experience levels. It divides participants into three groups: Group A: Uses CineVision, with professional directors (D1-D3) and amateur directors (AD1-AD9), along with professional cinematographers (C1-C3) and amateur cinematographers (AC1-AC9).
Group B: Uses AI tools, also categorized with professional and amateur participants. Group C: Uses traditional hand-drawn storyboards. The teams are formed by mixing professional and amateur participants from different groups, allowing for a variety of experiences and tool usage within each team. This structure enables a comparison of how different tools and levels of experience affect the outcomes of the storyboard creation process.
}
}
\end{table}

\subsection{Procedure}
Before the study began, the research protocol was approved by the Institutional Review Board (IRB): {HREP-2025-0092.} The study involved a 60-minute in-person experimental session. Participants first completed a background survey. Participants were divided into 3 groups. Upon arrival at the lab, Group A received a 10-minute system operation training, which focused on familiarizing them with the interface buttons of the system to ensure they understood the task requirements. The remaining two groups (B and C) of participants spent this time familiarizing themselves with the equipment they would be using—AI tools and hand-drawn storyboards. Apart from the creative tool, all participants worked with the same script and timeline to complete identical film storyboard creation tasks, ensuring consistency across the experimental design. Following the 10-minute training, participants had 50 minutes to complete the tasks outlined below:

\subsubsection{Storyboard Creation Task}
In this task, two participants (a director and a cinematographer) form a pair. Based on insights from previous expert interviews, it was found that during the pre-production phase, the process of translating a script into a storyboard is typically director-led, with the cinematographer supporting the director by offering suggestions and references. Therefore, this task requires each pair to collaborate on creating a storyboard based on a provided script, which consists of six consecutive dialogue scenes set in a single location. The lines of the script are as follows: 

1) Characters: 

Ethan: Male, 37, thoughtful, carrying hidden regret. 

Olivia: Female, 34, elegant, composed yet conflicted.

2) Dialogue: \textit{Ethan: After all these years, do you think we made the wrong choice?}

\textit{Olivia:  If I said no, I'd be lying—not just to you, but to myself.}

\textit{Ethan: Guess we've always been good at lying to ourselves.}

\textit{Olivia:  Some paths, once taken, don't lead back.}

\textit{Ethan:  But if you had the chance, would you risk it again?}

\textit{Olivia:  We can't afford another gamble, Ethan. Neither of us can.}

The participants must visually translate these dialogue scenes into a storyboard. They have 50 minutes to engage in discussions and creative collaboration, considering aspects such as reference material selection, light source choice, color, style, and actor costume design. The goal is to recreate the ideal effects of shadows, highlights, and overall atmosphere using appropriate lighting, thereby demonstrating their understanding of the script's content. Upon completion of the task, participants fill out the NASA-TLX and UEQ to evaluate their user experience. Additionally, each participant completed the Collaboration Score, a mutual 7-point rating that captured how well they believed their partner understood their shooting intentions (1 = completely misunderstood, 7 = completely understood). The task took place in a quiet conference room while researchers observed director–cinematographer communication in real time.

\subsection{Data Analysis Approach}
Following previous studies \cite{do2024stepping,wei2025illuminating}, we initially computed the mean and total scores for all subscale items. Subsequently, the Shapiro-Wilk test \cite{razali2011power} was conducted to evaluate the normality of the score distributions. Given that the results indicated non-normality for some of the variables, we performed a non-parametric Kruskal-Wallis test \cite{mckight2010kruskal} to compare the scores across the three experimental groups: Group A (\textit{CineVision}), Group B (AI tools), and Group C (traditional hand-drawn storyboards). To investigate significant differences between groups in more detail, post-hoc pairwise comparisons were conducted using Dunn's test \cite{dinno2015nonparametric}, with a Bonferroni correction applied to control for multiple comparisons.

\section{Results}
Table \ref{table result} illustrates the means and standard deviations for all measures. We begin by exploring the results of the NASA-TLX, then present the findings of the User Experience Questionnaire (UEQ) in turn, and finally summarize the conclusions obtained from the semi-structured interviews.

\begin{figure*}[t]
    \centering
    \includegraphics[width=17.5cm]{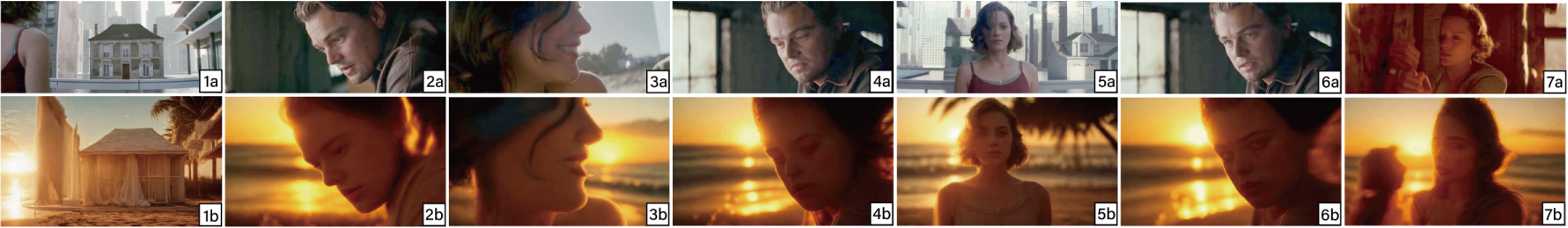}
    \caption{A sample from the Group A user study task results of the \textit{CineVision} system: (1a)-(7a) The director and cinematographer selected reference image through the system. (1b)-(7b) The storyboard generated after communication between the director and the cinematographer; (1b) Ambient sounds: sound of waves and seagulls chirping. (2b) Ethan: After all these years, do you think we made the wrong choice?. (3b) Olivia: If I said no, I'd be lying—not just to you, but to myself. (4b) Ethan: Guess we've always been good at lying to ourselves. (5b) Olivia:  Some paths, once taken, don't lead back. (6b) Ethan: But if you had the chance, would you risk it again? (7b) Olivia:  We can't afford another gamble, Ethan. Neither of us can.}
    \label{GroupA}
    \Description{A Sample from the Group A User Study Task Results of the \textit{CineVision} System: (1a)-(7a) The director and cinematographer selected reference image through the system. (1b)-(7b) The storyboard generated after communication between the director and the cinematographer; (1b) Ambient sounds: sound of waves and seagulls chirping. (2b) Ethan: After all these years, do you think we made the wrong choice?. (3b) Olivia: If I said no, I'd be lying—not just to you, but to myself. (4b) Ethan: Guess we've always been good at lying to ourselves. (5b) Olivia:  Some paths, once taken, don't lead back. (6b) Ethan: But if you had the chance, would you risk it again? (7b) Olivia:  We can't afford another gamble, Ethan. Neither of us can.}
\end{figure*}

\begin{figure*}[ht]
    \centering
    \includegraphics[width=17cm]{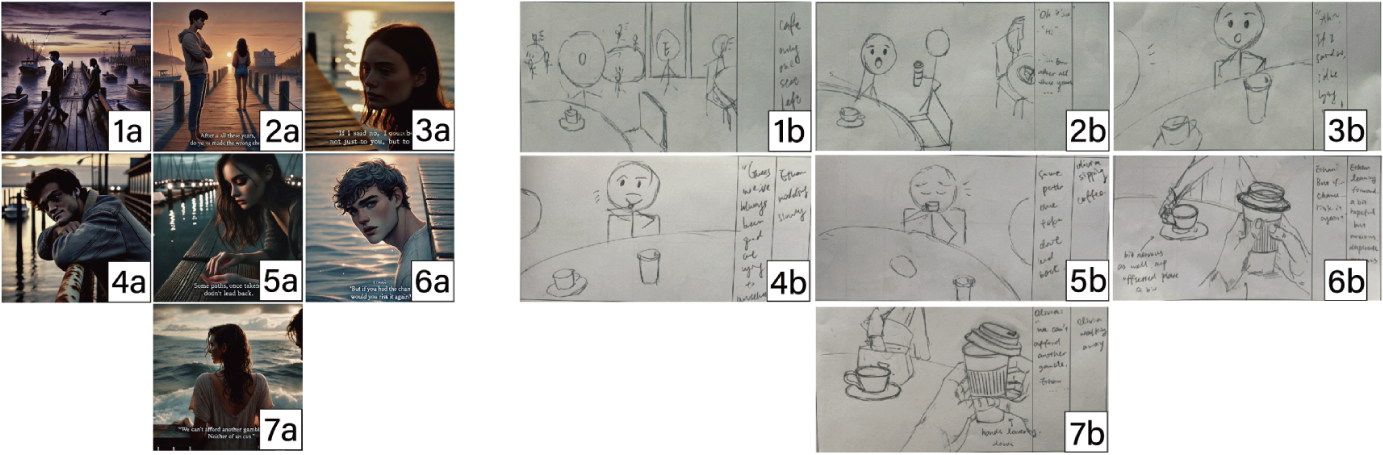}
    \caption{Two samples from the Group B and Group C user study task results: (1a)-(7a) The storyboard generated by \textit{DALL·E 3}  (Group B). (1b)-(7b) The storyboard hand-drawn  (Group C).} 
    \Description{Two Samples from the Group B and Group C User Study Task Results: (1a)-(7a) The storyboard generated by \textit{DALL·E 3} after communication between the director and the cinematographer (Group B). (1b)-(7b) The storyboard hand-drawn after communication between the director and the cinematographer (Group C). (2a/2b) Ethan: After all these years, do you think we made the wrong choice?. (3a/3b) Olivia:  If I said no, I'd be lying—not just to you, but to myself. (4a/4b) Ethan: Guess we've always been good at lying to ourselves. (5a/5b) Olivia: Some paths, once taken, don't lead back. (6a/6b) Ethan:  But if you had the chance, would you risk it again? (7a/7b) Olivia: We can't afford another gamble, Ethan. Neither of us can.}
    \label{GroupB}
\end{figure*}

\begin{table}[ht]
\caption{An overview of the mean (M) and standard deviation (SD) for the questionnaire results is provided. Asterisks (*) denote a significant difference between Group A and Group B, a plus sign (+) indicates a significant difference between Group A and Group C, and a hash symbol (\#) represents a significant difference between Group B and Group C. For NASA-TLX scores, a rating of 1 represents the best experience, while a rating of 7 reflects the worst. In contrast, for User Experience (UE) scores, a rating of 1 signifies the worst experience, and a rating of 7 signifies the best experience.}
\scalebox{1}{
\begin{tabular}{cccc}
\hline
                    & \begin{tabular}[c]{@{}c@{}}Group A\\ M (SD)\end{tabular} & \begin{tabular}[c]{@{}c@{}}Group B\\ M (SD)\end{tabular} & \begin{tabular}[c]{@{}c@{}}Group C\\ M (SD)\end{tabular} \\ \hline
NASA-Q              &                                                          &                                                          &                                                          \\ \hline
Physical Demand*+\# & 2.00 (1.31)                                               & 2.63 (1.30)                                               & 5.25 (1.28)                                               \\
Temporal Demand*+\# & 1.75 (1.04)                                               & 4.25(1.98)                                               & 4.75 (1.58)                                               \\
Effort*+\#          & 2.75 (1.64)                                               & 4.75(1.67)                                               & 5.25 (1.16)                                               \\
Frustration*+\#     & 2.25 (1.04)                                               & 3.13(1.96)                                               & 5.00 (1.60)                                               \\
Performance         & 3.13 (1.55)                                               & 2.50(0.53)                                               & 3.50 (1.85)                                               \\
Mental Demand       & 3.63 (1.18)                                               & 4.88(1.36)                                               & 5.00 (0.93)                                               \\ \hline
UE-Q                &                                                          &                                                          &                                                          \\ \hline
Usefulness*+\#      & 6.00 (0.93)                                               & 4.75 (2.05)                                               & 4.38 (0.74)                                               \\
Ease of Use*+\#     & 5.88 (1.25)                                               & 4.88 (1.96)                                               & 3.75 (1.16)                                               \\
Overall Experience  & 5.75 (0.71)                                               & 4.38 (1.85)                                               & 5.38 (1.51)                                               \\ \hline
\end{tabular}
\Description{This table provides an overview of the mean (M) and standard deviation (SD) of questionnaire results across three groups using different tools (CineVision, AI tools, and traditional hand-drawn storyboards). The data is segmented into NASA-TLX scores and User Experience (UE) scores. NASA-TLX Scores: This section measures the physical, temporal, mental, and effort demands, as well as frustration and performance. A higher rating (7) indicates a worse experience, while a lower rating (1) indicates the best experience. Asterisks (*) denote significant differences between Group A and Group B, plus signs (+) indicate significant differences between Group A and Group C, and hash symbols (#) show significant interaction effects between Group B and Group C. User Experience (UE) Scores: This section measures the perceived usefulness, ease of use, and overall experience. A lower rating (1) represents the worst experience, while a higher rating (7) signifies the best experience. The table reveals how the different groups felt about the tools based on these metrics. The results show that Group A (CineVision) generally had lower physical and temporal demands, while Group B (AI tools) experienced more frustration. Group C (traditional storyboards) had higher frustration and temporal demands but performed similarly in overall experience ratings compared to the other groups.}
\label{table result}
}
\end{table}

\subsection{NASA-TLX Questionnaire}
\begin{figure}[t]
    \centering
    \includegraphics[width=8.5cm]{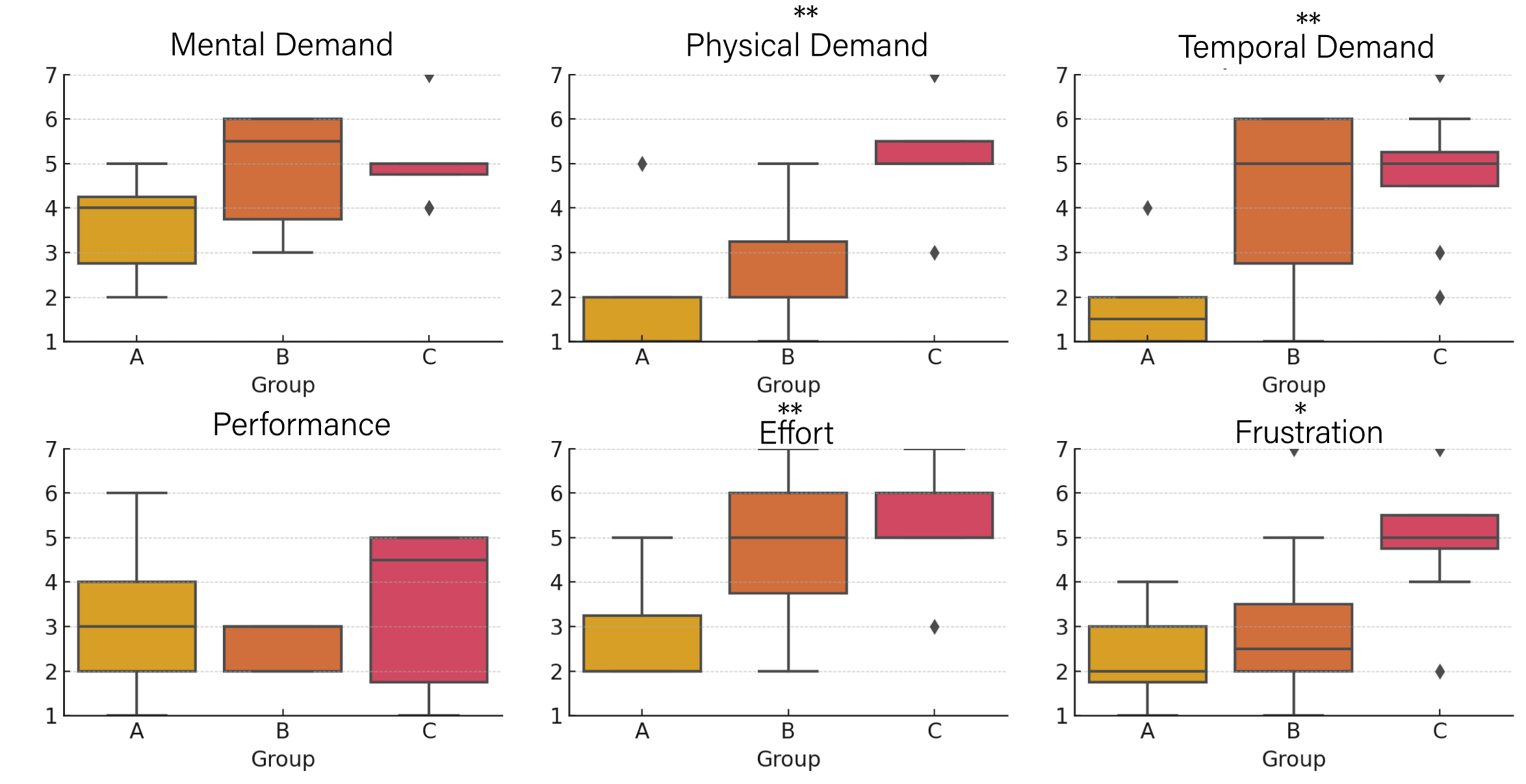}
    \caption{{Boxplots of NASA-TLX workload ratings across three groups (A, B, C) for six dimensions: Mental Demand, Physical Demand, Temporal Demand, Performance, Effort, and Frustration (* : p ~\textless 0.05, ** : p ~\textless 0.01).}}
    \label{nasa}
\end{figure}

Analysis of the NASA-TLX questionnaire data revealed clear differences in the patterns across the measurement variables. There were no significant differences between the three groups in terms of Performance and Mental Demand. However, significant differences were observed in Physical Demand, Temporal Demand, Effort, and Frustration, {as shown in Figure \ref{nasa}.}

Specifically, for Physical Demand, Group A (${M}$ = 2.00, SD = 1.31), Group B (${M}$ = 2.63, SD = 1.30), and Group C (${M}$ = 5.25, SD = 1.28) showed significant differences ($\chi^{2}(2) = 13.11$, p~\textless 0.01). Similarly, Temporal Demand showed significant differences ($\chi^{2}(2) = 10.29$, p~\textless 0.01), with Group A (${M}$ = 1.75, SD = 1.04) significantly differing from both Group B (${M}$ = 4.25, SD = 1.98) and Group C (${M}$ = 4.75, SD = 1.58). In terms of Effort, significant differences were found ($\chi^{2}(2) = 9.78$, p~\textless 0.01), with Group A (${M}$ = 2.75, SD = 1.64) differing significantly from both Group B (${M}$ = 4.75, SD = 1.67) and Group C (${M}$ = 5.25, SD = 1.16). In Frustration, significant differences were observed ($\chi^{2}(2) = 8.53$, p~\textless 0.05), with Group A (${M}$ = 2.25, SD = 1.04) significantly differing from both Group B (${M}$ = 3.13, SD = 1.96) and Group C (${M}$ = 5.00, SD = 1.60). These findings highlight significant differences in workload dimensions across the three groups.

\subsection{User Experience Questionnaire}
\begin{figure}[t]
    \centering
    \includegraphics[width=7.5cm]{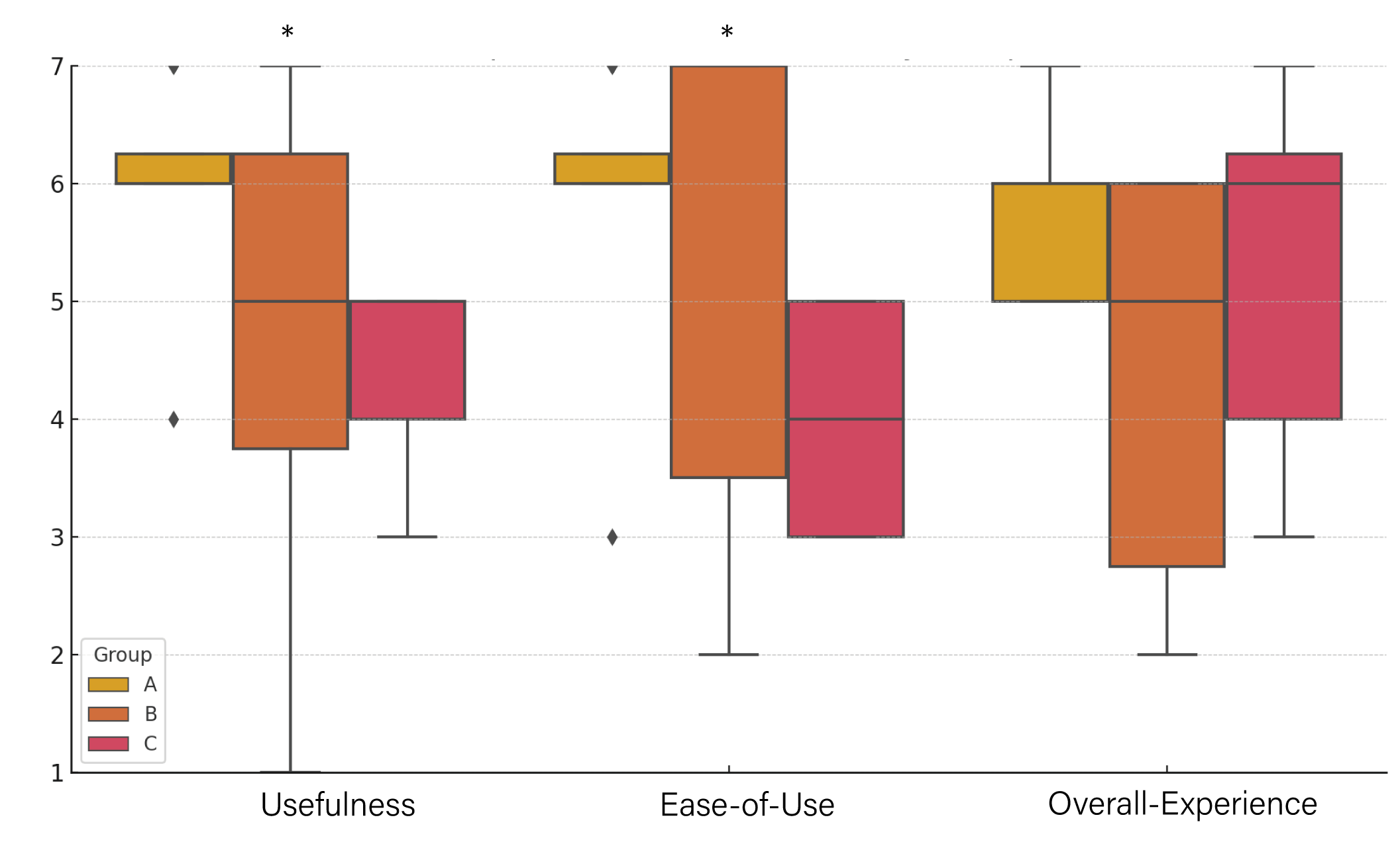}
    \caption{{Boxplots of UE-Q ratings across three groups (A, B, C) on three dimensions: Usefulness, Ease of Use, and Overall Experience (* : p ~\textless 0.05).}}
    \label{UEQ}
\end{figure}

The results for three conditions have three dimensions: ``Overall Experience'', ``Usefulness'', and ``Ease of Use''. Each dimension has one questionnaire, resulting in a total of three questions, {as shown in Figure \ref{UEQ}.}

Usefulness: For the Question (Q1) `{\textit{``This workflow approach is a time saver for me and can be done quickly.}}' showed significant differences ($\chi^{2}(2) = 6.88$, p ~\textless 0.05): Group A (${M}$ = 6.00, SD = 0.93), B (${M}$ = 4.75, SD = 2.05) and C (${M}$ = 4.38, SD = 0.74). Ease of Use: For the Question (Q2) `{\textit{``I can easily master this mode of creating storyboards.}}' showed significant differences ($\chi^{2}(2) = 6.55$, p ~\textless 0.05): Group A (${M}$ = 5.88, SD = 1.25), Group B (${M}$ = 4.88, SD = 1.96), and Group C (${M}$ = 3.75, SD = 1.16). Specifically, Group A reported a higher ease of use compared to Group B and Group C. Overall Experience: For the Question (Q3) `{\textit{I think creating storyboards using this method would work well.}}', Group A (${M}$ = 5.75, SD = 0.71), B (${M}$ = 4.38, SD = 1.85) and C (${M}$ = 5.38, SD = 1.51), showed no significant differences.

\subsection{Task Results}
In the ``Storyboard Creation'' task, we evaluated 24 participants—eight in each of three groups, with two participants per team—comprising both professionals and amateurs. Figure \ref{GroupA} shows the results produced with the \textit{CineVision} system, Figure \ref{GroupB} (1a)-(7a) illustrates outcomes generated using \textit{DALL·E 3}, and Figure \ref{GroupB} (1b)-(7b) displays hand-drawn outputs from the same task.

{The collaboration score} between directors and cinematographers showed significant differences ($\chi^{2}(2) = 9.16$, p~\textless 0.05). Group A, which used the \textit{CineVision} system, had the highest ratings (${M}$ = 6.5, SD = 0.53), followed by Group B, which used AI-assisted creation (\textit{DALL·E 3}) (${M}$ = 5.12, SD = 1.54), and Group C, which used hand-drawn storyboards (${M}$ = 4.13, SD = 1.46). {These results indicate a trend toward higher perceived mutual understanding, but further work is needed to confirm whether this translates into on-set efficiency or decision authority.}

\subsection{Observation and Semi-structured interviews}
In our observation of the participants, we noticed clear gender differences in their communication and collaboration methods. Specifically, for the participants in Group A, the system's real-time visualization led them to focus more on the overall style of the characters and environments, including how lighting, character expressions, and background elements conveyed emotional atmospheres. In contrast, Group B participants spent a significant amount of time refining feedback and regenerating AI-generated images. Due to the uncontrollable nature of AI outputs, they constantly discussed and adjusted the prompts to improve the results, requiring a great deal of time and effort. Group C, based on individual drawing skills, showed varying communication dynamics. Teams with stronger drawing skills communicated more effectively, while teams with limited drawing skills relied heavily on visual references and repeatedly modified sketches to create accurate storyboards. This collaboration style led directors and cinematographers to focus most of their energy on visual descriptions and translating text into visual representations, leaving little time or energy for more detailed adjustments in environment, lighting, and character design.

We also observed that Group A's task uniquely triggered extensive communication and interaction among participants. Most groups discussed the impact of background and lighting changes on reference images, repeatedly testing different effects to match the emotions conveyed in the script. They also explored combinations of director styles, seeking more stylized interpretations. Notably, in on-site discussions, directors and cinematographers often tested the modifications suggested by each other and discussed the reasons why one visual effect was better than another.

Our semi-structured interviews further revealed that participants highly appreciated \textit{CineVision}'s real-time feedback on visual modifications. As Participant C1 mentioned: \textit{``Compared to this system, traditional hand-drawing methods take much longer to create 7 shots. First, the visual precision of hand-drawn storyboards can't compare with this system, and second, the atmosphere, colors, and communication provided by the system are much more accurate. It saves us a lot of time in discussions with the director.''} Director D1 pointed out, \textit{``The system's biggest advantage is the large amount of visual references it provides, which greatly helps in selecting shots and angles.''} In terms of the creative process, some creators prefer working independently. AD4 (Group B) focused on providing a more complete story for cinematography based on the script, while AC4 (Group B, cinematographer) had to repeatedly adjust based on the director’s story to facilitate AI system usage. After the interview, AC4 also mentioned, \textit{``I feel like I've done too much of what should have been the director's job, and it felt like my workload was heavier.''} Additionally, Group C mentioned, \textit{``The accuracy of drawing is the biggest challenge in communication,''} as AD8 (Group C, director) pointed out. However, participants using \textit{CineVision} were more inclined to work collaboratively. The real-time visibility provided by the system encouraged them to share ideas and actively participate in modifications.

\section{DISCUSSION}
In this section, we examine \textit{CineVision}’s influence on the pre-production workflow. First, we show how real-time visualization speeds collaboration and reduces creative friction, followed by an exploration of how dynamic lighting fosters stronger consensus among directors and cinematographers. We then highlight the importance of character and costume customization for continuity, before discussing how AI-driven tools might reshape director–cinematographer collaboration. Finally, we address current limitations and propose directions for future enhancements.

\subsection{Instant Visualization for Collaboration}
Compared to the two baseline methods, \textit{CineVision}’s real-time visualization capability significantly reduces mental demands and increases iteration speed. This immediate feedback loop gives directors and cinematographers a dynamic visual reference, allowing them to develop or adjust scenes while effectively bridging the gap between script and imagery in real time. By instantly visualizing scenes, creative decisions, such as adjusting camera angles or blocking actors, can be evaluated on the spot, eliminating the prolonged guesswork typical of traditional workflows. From a creative standpoint, relying less on imagination and memory lowers the team’s overall mental effort. Directors no longer have to mentally piece together complex scenes or verbally describe every visual cue; instead, they can see a tangible depiction of their ideas and refine them iteratively. {The immediate visual feedback appeared to reduce misunderstandings in our lab setting; future field studies should verify whether similar gains occur in real productions.}

In contrast, traditional hand-drawn storyboards do not offer immediate updates: modifying a shot requires redrawing or extensive annotations, a time-consuming process that disrupts creative flow. Existing AI-based tools, while faster than hand drawing, often lack truly interactive adjustments; they typically require re-generating images for every modification, resulting in a fragmented and mentally demanding iteration process. Moreover, traditional storyboards are inherently static and detached from dynamic elements like camera movement or lighting, forcing filmmakers to rely on mental extrapolation when making changes. By seamlessly integrating writing and visualization, \textit{CineVision} circumvents these limitations, enabling continuous, real-time refinement. In our study, participants using \textit{CineVision} completed storyboarding tasks faster and with fewer modifications than those working with the baseline methods, underscoring the system’s efficiency advantage. They noted that the platform’s seamless updates allowed them to test creative ideas freely without losing momentum, whereas each change under the baseline methods introduced substantial delays. Consequently, real-time visualization  not only accelerates pre-production but also preserves creative momentum and clarity, {which participants perceived as advantageous over the two baselines in our study.}

\subsection{Lighting for Enhanced Visual Consensus}
Lighting is not merely about illumination—it shapes the emotional tone and narrative focus of a film \cite{tan2018psychology,alton2013painting}. Traditional hand-drawn storyboards struggle to convey real-time changes in light source, shadow, and ambiance; as a result, directors and cinematographers often rely on text notes or lengthy conversations to bridge these gaps, which can lead to misinterpretation and costly revisions. In contrast, \textit{CineVision}’s dynamic lighting control of \textit{ CineVision } allows teams to rapidly experiment with and compare various lighting setups, such as changing from a soft, high-key style to a moody, low-key environment, with immediate visual feedback. By using physically consistent relighting algorithms, \textit{CineVision} accurately simulates shadow behavior, highlights, and overall tonal shifts in a 2D context, approximating the fidelity of a 3D lighting setup without the overhead of building detailed 3D models. This capability not only accelerates creative iteration but also fosters stronger alignment among key stakeholders early in the process. Directors and cinematographers can lock in major lighting decisions, such as adding a strong backlight to create silhouettes or adjusting fill lights for softer shadows, long before production begins, mitigating risks of confusion or last-minute changes on set. In essence, \textit{CineVision}’s dynamic lighting transforms storyboards from static placeholders into interactive previews, {potentially refining the efficiency and clarity of pre-visualization, according to participant feedback.}

\subsection{Consistent Characters and Costumes}

\textit{CineVision}’s character and costume customization feature meets the critical need for continuity and visual specificity in storyboards. Once a user sets the appearance of a character (face, hairstyle, clothing and accessories), those attributes automatically carry over to every shot, ensuring immediate visual consistency for both the narrative and the shot transitions. Directors can also experiment with different outfits or props without having to redraw the entire scene. In user studies, participants noted that consistently seeing the same character features and costumes helped the team maintain a unified understanding of the character's portrayal, an advantage rarely achieved with hand-drawn boards. Furthermore, because characters are integrated with scene design, \textit{CineVision} takes into account costume visibility during lighting adjustments (for example, ensuring dark clothing remains visible in low-light scenes), letting directors preview how character attire interacts with on-set lighting. In contrast, manual storyboards require artists to repeatedly refer to models or design sheets to maintain character consistency, which is time-consuming and prone to visual drift. Likewise, the AI baseline tool we tested lacked a built-in mechanism for character continuity, often resulting in mismatched faces or outfits in different shots. \textit{CineVision}, however, preserves each character’s identity and attire from the moment it is defined, significantly reducing time spent clarifying ``who's who'' and allowing creators to focus on the core of the story.

\subsection{Evolving Director–Cinematographer Collaboration}

AI pre-production systems, represented by \textit{CineVision}, have the potential to reshape the creative collaboration between directors and cinematographers. Traditionally, cinematographers rely on rough storyboards or verbal descriptions to communicate their vision, which cinematographers interpret and refine. With \textit{CineVision}, directors can now generate real-time, high-quality visualizations that precisely communicate their intended shot composition, lighting, and mood. This setup may lower miscommunication during early iterations, though roles could shift as creative authority is redistributed. This dynamic shift empowers directors to take a more hands-on role in the visual pre-visualization process, lowering the cognitive load on cinematographers to interpret abstract visual cues. However, it also raises questions about the potential diminution of the cinematographer's creative role. By providing directors with a tool to directly manipulate visuals, there is a risk of reducing the cinematographer’s input to technical execution rather than creative collaboration. The balance between creative agency and technical execution may evolve, where the cinematographer’s expertise is channeled through real-time adjustments within the AI-driven system, rather than through traditional, more independent interpretations of the director’s intent. In the end, \textit{CineVision}'s integration into the pre-production process could lead to a more symbiotic collaboration if it is seen as a tool that enhances communication and creative flexibility rather than as a replacement for the cinematographer's interpretative work.

\subsection{Limitations and Future Work}
{Although \textit{CineVision} greatly improves early pre-visualization, our evaluation revealed several limitations that will guide its next iterations. First, to preserve real-time feedback, the system renders at a relatively low resolution; processing dense crowds or intricate patterns can slow performance and break creative flow. Second, while it copes well with conventional dialogue and moderate action, quality degrades in large-scale or highly layered scenes where overlapping elements obscure spatial relationships. Third, style customization depends on a fixed library—if a director’s desired look is absent, the system cannot reproduce it. We plan to let users upload reference material and, with minimal ML expertise, fine-tune bespoke styles. Fourth, collaboration is framed around a director–cinematographer dyad; other departments (production design, costume, lighting, VFX) can only observe rather than participate in real time. Fifth, \textit{CineVision} generates 2-D stills without shot-critical metadata (lens focal length, T-stop, camera height, sensor crop) or automatic continuity checks, so cinematographers must translate boards into shot lists and verify eye-lines, lens choices, and movement arcs by hand. Sixth, our study was limited to short, dyadic lab sessions; full-crew, longitudinal on-set studies with interaction logs and talk-back recordings are needed to validate communication benefits at production scale. These gaps reflect deliberate trade-offs: the current release optimizes speed for ideation, but offers diminishing returns for tasks demanding lens-accurate metadata, strict continuity management, or ACES-compliant color pipelines. We therefore view \textit{CineVision} as an incremental, staircase development: future versions will expose a tiered ``sketch'' versus ``production'' mode, add industry-standard exports (FBX/CSV, Unreal Engine), and support multi-role collaboration, ensuring that deeper controls appear only when they deliver clear professional value.}

\section{CONCLUSION}
\textit{CineVision} streamlines the pre-production workflow by merging real-time visualization, dynamic lighting control, and character continuity into one coherent platform. {Our preliminary findings indicate reduced task time and mental demand for the tested storyboard task; verifying long-term creative alignment warrants further study.} While opportunities remain to improve resolution, expand style customization, and enable broader multi-user collaboration, {our findings highlight \textit{CineVision}’s promise to support visual communication and decision-making in pre-production.}

\begin{acks}
We thank the anonymous reviewers for their constructive comments and suggestions. This work was partially supported by the HKUST BGF (BGF.2025.004) and the RGC TRS grant (T22-607/24-N). We are also grateful to Fanyu Meng for his contribution to video production.
\end{acks}

\bibliographystyle{ACM-Reference-Format}
\bibliography{UIST}


\begin{thebibliography}{92}


\ifx \showCODEN    \undefined \def \showCODEN     #1{\unskip}     \fi
\ifx \showDOI      \undefined \def \showDOI       #1{#1}\fi
\ifx \showISBNx    \undefined \def \showISBNx     #1{\unskip}     \fi
\ifx \showISBNxiii \undefined \def \showISBNxiii  #1{\unskip}     \fi
\ifx \showISSN     \undefined \def \showISSN      #1{\unskip}     \fi
\ifx \showLCCN     \undefined \def \showLCCN      #1{\unskip}     \fi
\ifx \shownote     \undefined \def \shownote      #1{#1}          \fi
\ifx \showarticletitle \undefined \def \showarticletitle #1{#1}   \fi
\ifx \showURL      \undefined \def \showURL       {\relax}        \fi
\providecommand\bibfield[2]{#2}
\providecommand\bibinfo[2]{#2}
\providecommand\natexlab[1]{#1}
\providecommand\showeprint[2][]{arXiv:#2}

\bibitem[Abid et~al\mbox{.}(2019)]%
        {abid2019gradio}
\bibfield{author}{\bibinfo{person}{Abubakar Abid}, \bibinfo{person}{Ali Abdalla}, \bibinfo{person}{Ali Abid}, \bibinfo{person}{Dawood Khan}, \bibinfo{person}{Abdulrahman Alfozan}, {and} \bibinfo{person}{James Zou}.} \bibinfo{year}{2019}\natexlab{}.
\newblock \showarticletitle{Gradio: Hassle-free sharing and testing of ML models in the wild}.
\newblock \bibinfo{journal}{\emph{arXiv preprint arXiv:1906.02569}} (\bibinfo{year}{2019}).
\newblock


\bibitem[Alton(2013)]%
        {alton2013painting}
\bibfield{author}{\bibinfo{person}{John Alton}.} \bibinfo{year}{2013}\natexlab{}.
\newblock \bibinfo{booktitle}{\emph{Painting with light}}.
\newblock \bibinfo{publisher}{University of California Press}.
\newblock


\bibitem[Bartindale et~al\mbox{.}(2012)]%
        {bartindale2012storycrate}
\bibfield{author}{\bibinfo{person}{Tom Bartindale}, \bibinfo{person}{Alia Sheikh}, \bibinfo{person}{Nick Taylor}, \bibinfo{person}{Peter Wright}, {and} \bibinfo{person}{Patrick Olivier}.} \bibinfo{year}{2012}\natexlab{}.
\newblock \showarticletitle{StoryCrate: tabletop storyboarding for live film production}. In \bibinfo{booktitle}{\emph{Proceedings of the SIGCHI Conference on Human Factors in Computing Systems}}. \bibinfo{pages}{169--178}.
\newblock


\bibitem[Bengesi et~al\mbox{.}(2024)]%
        {bengesi2024advancements}
\bibfield{author}{\bibinfo{person}{Staphord Bengesi}, \bibinfo{person}{Hoda El-Sayed}, \bibinfo{person}{Md~Kamruzzaman Sarker}, \bibinfo{person}{Yao Houkpati}, \bibinfo{person}{John Irungu}, {and} \bibinfo{person}{Timothy Oladunni}.} \bibinfo{year}{2024}\natexlab{}.
\newblock \showarticletitle{Advancements in Generative AI: A Comprehensive Review of GANs, GPT, Autoencoders, Diffusion Model, and Transformers.}
\newblock \bibinfo{journal}{\emph{IEEE Access}} (\bibinfo{year}{2024}).
\newblock


\bibitem[Biermann et~al\mbox{.}(2022)]%
        {biermann2022tool}
\bibfield{author}{\bibinfo{person}{Oloff~C Biermann}, \bibinfo{person}{Ning~F Ma}, {and} \bibinfo{person}{Dongwook Yoon}.} \bibinfo{year}{2022}\natexlab{}.
\newblock \showarticletitle{From tool to companion: Storywriters want AI writers to respect their personal values and writing strategies}. In \bibinfo{booktitle}{\emph{Proceedings of the ACM Designing Interactive Systems Conference}}. \bibinfo{pages}{1209--1227}.
\newblock


\bibitem[Block(2020)]%
        {block2020visual}
\bibfield{author}{\bibinfo{person}{Bruce Block}.} \bibinfo{year}{2020}\natexlab{}.
\newblock \bibinfo{booktitle}{\emph{The visual story: Creating the visual structure of film, TV, and digital media}}.
\newblock \bibinfo{publisher}{Routledge}.
\newblock


\bibitem[Boutros et~al\mbox{.}(2023)]%
        {boutros2023idiff}
\bibfield{author}{\bibinfo{person}{Fadi Boutros}, \bibinfo{person}{Jonas~Henry Grebe}, \bibinfo{person}{Arjan Kuijper}, {and} \bibinfo{person}{Naser Damer}.} \bibinfo{year}{2023}\natexlab{}.
\newblock \showarticletitle{Idiff-face: Synthetic-based face recognition through fizzy identity-conditioned diffusion model}. In \bibinfo{booktitle}{\emph{Proceedings of the IEEE/CVF International Conference on Computer Vision}}. \bibinfo{pages}{19650--19661}.
\newblock


\bibitem[Brewster and Shafer(2011)]%
        {brewster2011fundamentals}
\bibfield{author}{\bibinfo{person}{Karen Brewster} {and} \bibinfo{person}{Melissa Shafer}.} \bibinfo{year}{2011}\natexlab{}.
\newblock \bibinfo{booktitle}{\emph{Fundamentals of theatrical design: A guide to the basics of scenic, costume, and lighting design}}.
\newblock \bibinfo{publisher}{Skyhorse Publishing Inc.}
\newblock


\bibitem[Brown(2016)]%
        {brown2016cinematography}
\bibfield{author}{\bibinfo{person}{Blain Brown}.} \bibinfo{year}{2016}\natexlab{}.
\newblock \bibinfo{booktitle}{\emph{Cinematography: theory and practice: image making for cinematographers and directors}}.
\newblock \bibinfo{publisher}{Routledge}.
\newblock


\bibitem[Chen et~al\mbox{.}(2023)]%
        {chen2023survey}
\bibfield{author}{\bibinfo{person}{Hung-Jen Chen}, \bibinfo{person}{Hong-Han Shuai}, {and} \bibinfo{person}{Wen-Huang Cheng}.} \bibinfo{year}{2023}\natexlab{}.
\newblock \showarticletitle{A survey of artificial intelligence in fashion}.
\newblock \bibinfo{journal}{\emph{IEEE Signal Processing Magazine}} \bibinfo{volume}{40}, \bibinfo{number}{3} (\bibinfo{year}{2023}), \bibinfo{pages}{64--73}.
\newblock


\bibitem[Chung and Adar(2023)]%
        {chung2023artinter}
\bibfield{author}{\bibinfo{person}{John Joon~Young Chung} {and} \bibinfo{person}{Eytan Adar}.} \bibinfo{year}{2023}\natexlab{}.
\newblock \showarticletitle{Artinter: AI-powered Boundary Objects for Commissioning Visual Arts}. In \bibinfo{booktitle}{\emph{Proceedings of the 2023 ACM Designing Interactive Systems Conference}}. \bibinfo{pages}{1997--2018}.
\newblock


\bibitem[Clark et~al\mbox{.}(2018)]%
        {clark2018creative}
\bibfield{author}{\bibinfo{person}{Elizabeth Clark}, \bibinfo{person}{Anne~Spencer Ross}, \bibinfo{person}{Chenhao Tan}, \bibinfo{person}{Yangfeng Ji}, {and} \bibinfo{person}{Noah~A Smith}.} \bibinfo{year}{2018}\natexlab{}.
\newblock \showarticletitle{Creative writing with a machine in the loop: Case studies on slogans and stories}. In \bibinfo{booktitle}{\emph{Proceedings of the 23rd International Conference on Intelligent User Interfaces}}. \bibinfo{pages}{329--340}.
\newblock


\bibitem[Concepcion(2022)]%
        {concepcion2022adobe}
\bibfield{author}{\bibinfo{person}{Rafael Concepcion}.} \bibinfo{year}{2022}\natexlab{}.
\newblock \bibinfo{booktitle}{\emph{Adobe Photoshop and Lightroom Classic for Photographers Classroom in a Book}}.
\newblock \bibinfo{publisher}{Adobe Press}.
\newblock


\bibitem[Croitoru et~al\mbox{.}(2023)]%
        {croitoru2023diffusion}
\bibfield{author}{\bibinfo{person}{Florinel-Alin Croitoru}, \bibinfo{person}{Vlad Hondru}, \bibinfo{person}{Radu~Tudor Ionescu}, {and} \bibinfo{person}{Mubarak Shah}.} \bibinfo{year}{2023}\natexlab{}.
\newblock \showarticletitle{Diffusion models in vision: A survey}.
\newblock \bibinfo{journal}{\emph{IEEE Transactions on Pattern Analysis and Machine Intelligence}} \bibinfo{volume}{45}, \bibinfo{number}{9} (\bibinfo{year}{2023}), \bibinfo{pages}{10850--10869}.
\newblock


\bibitem[Ding et~al\mbox{.}(2023)]%
        {ding2023diffusionrig}
\bibfield{author}{\bibinfo{person}{Zheng Ding}, \bibinfo{person}{Xuaner Zhang}, \bibinfo{person}{Zhihao Xia}, \bibinfo{person}{Lars Jebe}, \bibinfo{person}{Zhuowen Tu}, {and} \bibinfo{person}{Xiuming Zhang}.} \bibinfo{year}{2023}\natexlab{}.
\newblock \showarticletitle{Diffusionrig: Learning personalized priors for facial appearance editing}. In \bibinfo{booktitle}{\emph{Proceedings of the IEEE/CVF Conference on Computer Vision and Pattern Recognition}}. \bibinfo{pages}{12736--12746}.
\newblock


\bibitem[Dinno(2015)]%
        {dinno2015nonparametric}
\bibfield{author}{\bibinfo{person}{Alexis Dinno}.} \bibinfo{year}{2015}\natexlab{}.
\newblock \showarticletitle{Nonparametric pairwise multiple comparisons in independent groups using Dunn's test}.
\newblock \bibinfo{journal}{\emph{The Stata Journal}} \bibinfo{volume}{15}, \bibinfo{number}{1} (\bibinfo{year}{2015}), \bibinfo{pages}{292--300}.
\newblock


\bibitem[Do et~al\mbox{.}(2024)]%
        {do2024stepping}
\bibfield{author}{\bibinfo{person}{Tiffany~D Do}, \bibinfo{person}{Camille~Isabella Protko}, {and} \bibinfo{person}{Ryan~P McMahan}.} \bibinfo{year}{2024}\natexlab{}.
\newblock \showarticletitle{Stepping into the Right Shoes: The Effects of User-Matched Avatar Ethnicity and Gender on Sense of Embodiment in Virtual Reality}.
\newblock \bibinfo{journal}{\emph{IEEE Transactions on Visualization and Computer Graphics}} (\bibinfo{year}{2024}).
\newblock


\bibitem[Doust-Haghighi(2023)]%
        {doust2023laboratory}
\bibfield{author}{\bibinfo{person}{Elham Doust-Haghighi}.} \bibinfo{year}{2023}\natexlab{}.
\newblock \bibinfo{booktitle}{\emph{Laboratory Animation Productions: Strategies to Produce Customizable Animations to Be Used as Materials for Experimental Research on Media}}.
\newblock \bibinfo{publisher}{The University of Texas at Dallas}.
\newblock


\bibitem[Fan et~al\mbox{.}(2018)]%
        {fan2018hierarchical}
\bibfield{author}{\bibinfo{person}{Angela Fan}, \bibinfo{person}{Mike Lewis}, {and} \bibinfo{person}{Yann Dauphin}.} \bibinfo{year}{2018}\natexlab{}.
\newblock \showarticletitle{Hierarchical neural story generation}.
\newblock \bibinfo{journal}{\emph{arXiv preprint arXiv:1805.04833}} (\bibinfo{year}{2018}).
\newblock


\bibitem[Gal et~al\mbox{.}(2023)]%
        {gal2023encoder}
\bibfield{author}{\bibinfo{person}{Rinon Gal}, \bibinfo{person}{Moab Arar}, \bibinfo{person}{Yuval Atzmon}, \bibinfo{person}{Amit~H Bermano}, \bibinfo{person}{Gal Chechik}, {and} \bibinfo{person}{Daniel Cohen-Or}.} \bibinfo{year}{2023}\natexlab{}.
\newblock \showarticletitle{Encoder-based domain tuning for fast personalization of text-to-image models}.
\newblock \bibinfo{journal}{\emph{ACM Transactions on Graphics}} \bibinfo{volume}{42}, \bibinfo{number}{4} (\bibinfo{year}{2023}), \bibinfo{pages}{1--13}.
\newblock


\bibitem[Glebas(2012)]%
        {glebas2012directing}
\bibfield{author}{\bibinfo{person}{Francis Glebas}.} \bibinfo{year}{2012}\natexlab{}.
\newblock \bibinfo{booktitle}{\emph{Directing the story: professional storytelling and storyboarding techniques for live action and animation}}.
\newblock \bibinfo{publisher}{Routledge}.
\newblock


\bibitem[Graham(2022)]%
        {graham2022play}
\bibfield{author}{\bibinfo{person}{Katherine Graham}.} \bibinfo{year}{2022}\natexlab{}.
\newblock \showarticletitle{The play of light: rethinking mood lighting in performance}.
\newblock \bibinfo{journal}{\emph{Studies in Theatre and Performance}} \bibinfo{volume}{42}, \bibinfo{number}{2} (\bibinfo{year}{2022}), \bibinfo{pages}{139--155}.
\newblock


\bibitem[Guo et~al\mbox{.}(2023)]%
        {guo2023ai}
\bibfield{author}{\bibinfo{person}{Ziyue Guo}, \bibinfo{person}{Zongyang Zhu}, \bibinfo{person}{Yizhi Li}, \bibinfo{person}{Shidong Cao}, \bibinfo{person}{Hangyue Chen}, {and} \bibinfo{person}{Gaoang Wang}.} \bibinfo{year}{2023}\natexlab{}.
\newblock \showarticletitle{AI assisted fashion design: A review}.
\newblock \bibinfo{journal}{\emph{IEEE Access}}  \bibinfo{volume}{11} (\bibinfo{year}{2023}), \bibinfo{pages}{88403--88415}.
\newblock


\bibitem[Hart(2013)]%
        {hart2013art}
\bibfield{author}{\bibinfo{person}{John Hart}.} \bibinfo{year}{2013}\natexlab{}.
\newblock \bibinfo{booktitle}{\emph{The Art of the Storyboard: A filmmaker's introduction}}.
\newblock \bibinfo{publisher}{Routledge}.
\newblock


\bibitem[Hart(2006)]%
        {hart2006nasa}
\bibfield{author}{\bibinfo{person}{Sandra~G Hart}.} \bibinfo{year}{2006}\natexlab{}.
\newblock \showarticletitle{NASA-task load index (NASA-TLX); 20 years later}. In \bibinfo{booktitle}{\emph{Proceedings of the human factors and ergonomics society annual meeting}}, Vol.~\bibinfo{volume}{50}. \bibinfo{pages}{904--908}.
\newblock


\bibitem[He et~al\mbox{.}(2024a)]%
        {he2024cameractrl}
\bibfield{author}{\bibinfo{person}{Hao He}, \bibinfo{person}{Yinghao Xu}, \bibinfo{person}{Yuwei Guo}, \bibinfo{person}{Gordon Wetzstein}, \bibinfo{person}{Bo Dai}, \bibinfo{person}{Hongsheng Li}, {and} \bibinfo{person}{Ceyuan Yang}.} \bibinfo{year}{2024}\natexlab{a}.
\newblock \showarticletitle{Cameractrl: Enabling camera control for text-to-video generation}.
\newblock \bibinfo{journal}{\emph{arXiv preprint arXiv:2404.02101}} (\bibinfo{year}{2024}).
\newblock


\bibitem[He et~al\mbox{.}(2024b)]%
        {he2024dresscode}
\bibfield{author}{\bibinfo{person}{Kai He}, \bibinfo{person}{Kaixin Yao}, \bibinfo{person}{Qixuan Zhang}, \bibinfo{person}{Jingyi Yu}, \bibinfo{person}{Lingjie Liu}, {and} \bibinfo{person}{Lan Xu}.} \bibinfo{year}{2024}\natexlab{b}.
\newblock \showarticletitle{Dresscode: Autoregressively sewing and generating garments from text guidance}.
\newblock \bibinfo{journal}{\emph{ACM Transactions on Graphics}} \bibinfo{volume}{43}, \bibinfo{number}{4} (\bibinfo{year}{2024}), \bibinfo{pages}{1--13}.
\newblock


\bibitem[Hong et~al\mbox{.}(2023)]%
        {hong2023visual}
\bibfield{author}{\bibinfo{person}{Xudong Hong}, \bibinfo{person}{Asad Sayeed}, \bibinfo{person}{Khushboo Mehra}, \bibinfo{person}{Vera Demberg}, {and} \bibinfo{person}{Bernt Schiele}.} \bibinfo{year}{2023}\natexlab{}.
\newblock \showarticletitle{Visual writing prompts: Character-grounded story generation with curated image sequences}.
\newblock \bibinfo{journal}{\emph{Transactions of the Association for Computational Linguistics}}  \bibinfo{volume}{11} (\bibinfo{year}{2023}), \bibinfo{pages}{565--581}.
\newblock


\bibitem[Hosseini(2024)]%
        {hosseini2024generative}
\bibfield{author}{\bibinfo{person}{Donnesh~Dustin Hosseini}.} \bibinfo{year}{2024}\natexlab{}.
\newblock \bibinfo{title}{Generative AI: a problematic illustration of the intersections of racialized gender, race, ethnicity}.
\newblock
\newblock


\bibitem[Hou et~al\mbox{.}(2021)]%
        {hou2021towards}
\bibfield{author}{\bibinfo{person}{Andrew Hou}, \bibinfo{person}{Ze Zhang}, \bibinfo{person}{Michel Sarkis}, \bibinfo{person}{Ning Bi}, \bibinfo{person}{Yiying Tong}, {and} \bibinfo{person}{Xiaoming Liu}.} \bibinfo{year}{2021}\natexlab{}.
\newblock \showarticletitle{Towards high fidelity face relighting with realistic shadows}. In \bibinfo{booktitle}{\emph{Proceedings of the IEEE/CVF conference on computer vision and pattern recognition}}. \bibinfo{pages}{14719--14728}.
\newblock


\bibitem[Hu(2024)]%
        {hu2024animate}
\bibfield{author}{\bibinfo{person}{Li Hu}.} \bibinfo{year}{2024}\natexlab{}.
\newblock \showarticletitle{Animate anyone: Consistent and controllable image-to-video synthesis for character animation}. In \bibinfo{booktitle}{\emph{Proceedings of the IEEE/CVF Conference on Computer Vision and Pattern Recognition}}. \bibinfo{pages}{8153--8163}.
\newblock


\bibitem[Huang et~al\mbox{.}(2020)]%
        {huang2020movienet}
\bibfield{author}{\bibinfo{person}{Qingqiu Huang}, \bibinfo{person}{Yu Xiong}, \bibinfo{person}{Anyi Rao}, \bibinfo{person}{Jiaze Wang}, {and} \bibinfo{person}{Dahua Lin}.} \bibinfo{year}{2020}\natexlab{}.
\newblock \showarticletitle{Movienet: A holistic dataset for movie understanding}. In \bibinfo{booktitle}{\emph{Computer Vision--ECCV 2020: 16th European Conference, Glasgow, UK, August 23--28, 2020, Proceedings, Part IV 16}}. Springer, \bibinfo{pages}{709--727}.
\newblock


\bibitem[Ippolito et~al\mbox{.}(2019)]%
        {ippolito2019unsupervised}
\bibfield{author}{\bibinfo{person}{Daphne Ippolito}, \bibinfo{person}{David Grangier}, \bibinfo{person}{Chris Callison-Burch}, {and} \bibinfo{person}{Douglas Eck}.} \bibinfo{year}{2019}\natexlab{}.
\newblock \showarticletitle{Unsupervised hierarchical story infilling}. In \bibinfo{booktitle}{\emph{Proceedings of the First Workshop on Narrative Understanding}}. \bibinfo{pages}{37--43}.
\newblock


\bibitem[Jamison(2003)]%
        {jamison2003structured}
\bibfield{author}{\bibinfo{person}{D~Curtis Jamison}.} \bibinfo{year}{2003}\natexlab{}.
\newblock \showarticletitle{Structured query language (SQL) fundamentals}.
\newblock \bibinfo{journal}{\emph{Current protocols in bioinformatics}} \bibinfo{number}{1} (\bibinfo{year}{2003}), \bibinfo{pages}{9--2}.
\newblock


\bibitem[Katz(1991)]%
        {katz1991film}
\bibfield{author}{\bibinfo{person}{Steven~Douglas Katz}.} \bibinfo{year}{1991}\natexlab{}.
\newblock \bibinfo{booktitle}{\emph{Film directing shot by shot: visualizing from concept to screen}}.
\newblock \bibinfo{publisher}{Gulf Professional Publishing}.
\newblock


\bibitem[Keating(2009)]%
        {keating2009hollywood}
\bibfield{author}{\bibinfo{person}{Patrick Keating}.} \bibinfo{year}{2009}\natexlab{}.
\newblock \bibinfo{booktitle}{\emph{Hollywood Lighting from the Silent Era to Film Noir}}.
\newblock \bibinfo{publisher}{Columbia University Press}.
\newblock


\bibitem[Kim et~al\mbox{.}(2023)]%
        {kim2023dcface}
\bibfield{author}{\bibinfo{person}{Minchul Kim}, \bibinfo{person}{Feng Liu}, \bibinfo{person}{Anil Jain}, {and} \bibinfo{person}{Xiaoming Liu}.} \bibinfo{year}{2023}\natexlab{}.
\newblock \showarticletitle{Dcface: Synthetic face generation with dual condition diffusion model}. In \bibinfo{booktitle}{\emph{Proceedings of the IEEE/CVF Conference on Computer Vision and Pattern Recognition}}. \bibinfo{pages}{12715--12725}.
\newblock


\bibitem[Kim et~al\mbox{.}(2024)]%
        {kim2024gala}
\bibfield{author}{\bibinfo{person}{Taeksoo Kim}, \bibinfo{person}{Byungjun Kim}, \bibinfo{person}{Shunsuke Saito}, {and} \bibinfo{person}{Hanbyul Joo}.} \bibinfo{year}{2024}\natexlab{}.
\newblock \showarticletitle{Gala: Generating animatable layered assets from a single scan}. In \bibinfo{booktitle}{\emph{Proceedings of the IEEE/CVF Conference on Computer Vision and Pattern Recognition}}. \bibinfo{pages}{1535--1545}.
\newblock


\bibitem[Ko et~al\mbox{.}(2022)]%
        {ko2022we}
\bibfield{author}{\bibinfo{person}{Hyung-Kwon Ko}, \bibinfo{person}{Subin An}, \bibinfo{person}{Gwanmo Park}, \bibinfo{person}{Seung~Kwon Kim}, \bibinfo{person}{Daesik Kim}, \bibinfo{person}{Bohyoung Kim}, \bibinfo{person}{Jaemin Jo}, {and} \bibinfo{person}{Jinwook Seo}.} \bibinfo{year}{2022}\natexlab{}.
\newblock \showarticletitle{We-toon: A Communication Support System between Writers and Artists in Collaborative Webtoon Sketch Revision}. In \bibinfo{booktitle}{\emph{Proceedings of the 35th Annual ACM Symposium on User Interface Software and Technology}}. \bibinfo{pages}{1--14}.
\newblock


\bibitem[Li et~al\mbox{.}(2024)]%
        {li2024words}
\bibfield{author}{\bibinfo{person}{Danrui Li}, \bibinfo{person}{Samuel~S Sohn}, \bibinfo{person}{Sen Zhang}, \bibinfo{person}{Che-Jui Chang}, {and} \bibinfo{person}{Mubbasir Kapadia}.} \bibinfo{year}{2024}\natexlab{}.
\newblock \showarticletitle{From Words to Worlds: Transforming One-line Prompts into Multi-modal Digital Stories with LLM Agents}. In \bibinfo{booktitle}{\emph{Proceedings of the 17th ACM SIGGRAPH Conference on Motion, Interaction, and Games}}. \bibinfo{pages}{1--12}.
\newblock


\bibitem[Li et~al\mbox{.}(2022a)]%
        {li2022blip}
\bibfield{author}{\bibinfo{person}{Junnan Li}, \bibinfo{person}{Dongxu Li}, \bibinfo{person}{Caiming Xiong}, {and} \bibinfo{person}{Steven Hoi}.} \bibinfo{year}{2022}\natexlab{a}.
\newblock \showarticletitle{Blip: Bootstrapping language-image pre-training for unified vision-language understanding and generation}. In \bibinfo{booktitle}{\emph{International Conference on Machine Learning}}. PMLR, \bibinfo{pages}{12888--12900}.
\newblock


\bibitem[Li et~al\mbox{.}(2022b)]%
        {li2022physically}
\bibfield{author}{\bibinfo{person}{Zhengqin Li}, \bibinfo{person}{Jia Shi}, \bibinfo{person}{Sai Bi}, \bibinfo{person}{Rui Zhu}, \bibinfo{person}{Kalyan Sunkavalli}, \bibinfo{person}{Milo{\v{s}} Ha{\v{s}}an}, \bibinfo{person}{Zexiang Xu}, \bibinfo{person}{Ravi Ramamoorthi}, {and} \bibinfo{person}{Manmohan Chandraker}.} \bibinfo{year}{2022}\natexlab{b}.
\newblock \showarticletitle{Physically-based editing of indoor scene lighting from a single image}. In \bibinfo{booktitle}{\emph{European Conference on Computer Vision}}. Springer, \bibinfo{pages}{555--572}.
\newblock


\bibitem[Liu et~al\mbox{.}(2021)]%
        {liu2021generative}
\bibfield{author}{\bibinfo{person}{Ming-Yu Liu}, \bibinfo{person}{Xun Huang}, \bibinfo{person}{Jiahui Yu}, \bibinfo{person}{Ting-Chun Wang}, {and} \bibinfo{person}{Arun Mallya}.} \bibinfo{year}{2021}\natexlab{}.
\newblock \showarticletitle{Generative adversarial networks for image and video synthesis: Algorithms and applications}.
\newblock \bibinfo{journal}{\emph{Proc. IEEE}} \bibinfo{volume}{109}, \bibinfo{number}{5} (\bibinfo{year}{2021}), \bibinfo{pages}{839--862}.
\newblock


\bibitem[Liu et~al\mbox{.}(2023)]%
        {liu2023instaflow}
\bibfield{author}{\bibinfo{person}{Xingchao Liu}, \bibinfo{person}{Xiwen Zhang}, {and} \bibinfo{person}{Jianzhu et~al. Ma}.} \bibinfo{year}{2023}\natexlab{}.
\newblock \showarticletitle{Instaflow: One step is enough for high-quality diffusion-based text-to-image generation}. In \bibinfo{booktitle}{\emph{The Twelfth International Conference on Learning Representations}}.
\newblock


\bibitem[LoBrutto(2002)]%
        {lobrutto2002filmmaker}
\bibfield{author}{\bibinfo{person}{Vincent LoBrutto}.} \bibinfo{year}{2002}\natexlab{}.
\newblock \bibinfo{booktitle}{\emph{The filmmaker's guide to production design}}.
\newblock \bibinfo{publisher}{Simon and Schuster}.
\newblock


\bibitem[Lu et~al\mbox{.}(2023)]%
        {lu2023tf}
\bibfield{author}{\bibinfo{person}{Shilin Lu}, \bibinfo{person}{Yanzhu Liu}, {and} \bibinfo{person}{Adams Wai-Kin Kong}.} \bibinfo{year}{2023}\natexlab{}.
\newblock \showarticletitle{Tf-icon: Diffusion-based training-free cross-domain image composition}. In \bibinfo{booktitle}{\emph{Proceedings of the IEEE/CVF International Conference on Computer Vision}}. \bibinfo{pages}{2294--2305}.
\newblock


\bibitem[Lu et~al\mbox{.}(2024a)]%
        {lu2024mace}
\bibfield{author}{\bibinfo{person}{Shilin Lu}, \bibinfo{person}{Zilan Wang}, \bibinfo{person}{Leyang Li}, \bibinfo{person}{Yanzhu Liu}, {and} \bibinfo{person}{Adams Wai-Kin Kong}.} \bibinfo{year}{2024}\natexlab{a}.
\newblock \showarticletitle{Mace: Mass concept erasure in diffusion models}. In \bibinfo{booktitle}{\emph{Proceedings of the IEEE/CVF Conference on Computer Vision and Pattern Recognition}}. \bibinfo{pages}{6430--6440}.
\newblock


\bibitem[Lu et~al\mbox{.}(2024b)]%
        {lu2024robust}
\bibfield{author}{\bibinfo{person}{Shilin Lu}, \bibinfo{person}{Zihan Zhou}, \bibinfo{person}{Jiayou Lu}, \bibinfo{person}{Yuanzhi Zhu}, {and} \bibinfo{person}{Adams Wai-Kin Kong}.} \bibinfo{year}{2024}\natexlab{b}.
\newblock \showarticletitle{Robust watermarking using generative priors against image editing: From benchmarking to advances}.
\newblock \bibinfo{journal}{\emph{arXiv preprint arXiv:2410.18775}} (\bibinfo{year}{2024}).
\newblock


\bibitem[Matbouly(2022)]%
        {matbouly2022quantifying}
\bibfield{author}{\bibinfo{person}{Mustafa~Yousry Matbouly}.} \bibinfo{year}{2022}\natexlab{}.
\newblock \showarticletitle{Quantifying the unquantifiable: the color of cinematic lighting and its effect on audience’s impressions towards the appearance of film characters}.
\newblock \bibinfo{journal}{\emph{Current Psychology}} \bibinfo{volume}{41}, \bibinfo{number}{6} (\bibinfo{year}{2022}), \bibinfo{pages}{3694--3715}.
\newblock


\bibitem[Mateer(2017)]%
        {mateer2017directing}
\bibfield{author}{\bibinfo{person}{John Mateer}.} \bibinfo{year}{2017}\natexlab{}.
\newblock \showarticletitle{Directing for cinematic virtual reality: how the traditional film director’s craft applies to immersive environments and notions of presence}.
\newblock \bibinfo{journal}{\emph{Journal of Media Practice}} \bibinfo{volume}{18}, \bibinfo{number}{1} (\bibinfo{year}{2017}), \bibinfo{pages}{14--25}.
\newblock


\bibitem[McKight and Najab(2010)]%
        {mckight2010kruskal}
\bibfield{author}{\bibinfo{person}{Patrick~E McKight} {and} \bibinfo{person}{Julius Najab}.} \bibinfo{year}{2010}\natexlab{}.
\newblock \showarticletitle{Kruskal-wallis test}.
\newblock \bibinfo{journal}{\emph{The Corsini Encyclopedia of Psychology}} (\bibinfo{year}{2010}), \bibinfo{pages}{1}.
\newblock


\bibitem[Melki et~al\mbox{.}(2019)]%
        {melki2019investigation}
\bibfield{author}{\bibinfo{person}{Henry Melki}, \bibinfo{person}{Ian Montgomery}, {and} \bibinfo{person}{Greg Maguire}.} \bibinfo{year}{2019}\natexlab{}.
\newblock \emph{\bibinfo{title}{An Investigation into the Creative Processes in Generating Believable Photorealistic Film Characters}}.
\newblock \bibinfo{thesistype}{Ph.\,D. Dissertation}. \bibinfo{school}{Ulster University}.
\newblock


\bibitem[Mirowski et~al\mbox{.}(2023)]%
        {mirowski2023co}
\bibfield{author}{\bibinfo{person}{Piotr Mirowski}, \bibinfo{person}{Kory~W Mathewson}, \bibinfo{person}{Jaylen Pittman}, {and} \bibinfo{person}{Richard Evans}.} \bibinfo{year}{2023}\natexlab{}.
\newblock \showarticletitle{Co-writing screenplays and theatre scripts with language models: Evaluation by industry professionals}. In \bibinfo{booktitle}{\emph{Proceedings of the 2023 CHI Conference on Human Factors in Computing Systems}}. \bibinfo{pages}{1--34}.
\newblock


\bibitem[Painguzhali et~al\mbox{.}(2025)]%
        {painguzhali2025artificial}
\bibfield{author}{\bibinfo{person}{G Painguzhali}, \bibinfo{person}{Viswanath Ananth}, \bibinfo{person}{Kavitha}, {et~al\mbox{.}}} \bibinfo{year}{2025}\natexlab{}.
\newblock \showarticletitle{How artificial intelligence and generative AI is revolutionizing the fashion industry}.
\newblock In \bibinfo{booktitle}{\emph{Generative AI for Business Analytics and Strategic Decision Making in Service Industry}}. \bibinfo{publisher}{IGI Global Scientific Publishing}, \bibinfo{pages}{281--316}.
\newblock


\bibitem[Pal et~al\mbox{.}(2025)]%
        {pal2025illuminating}
\bibfield{author}{\bibinfo{person}{Arjama Pal}, \bibinfo{person}{Sakalya Mitra}, {and} \bibinfo{person}{D Lakshmi}.} \bibinfo{year}{2025}\natexlab{}.
\newblock \showarticletitle{Illuminating the path from script to screen using lights, camera, and AI}.
\newblock In \bibinfo{booktitle}{\emph{Transforming Cinema with Artificial Intelligence}}. \bibinfo{publisher}{IGI Global Scientific Publishing}, \bibinfo{pages}{97--142}.
\newblock


\bibitem[Pandey et~al\mbox{.}(2021)]%
        {pandey2021total}
\bibfield{author}{\bibinfo{person}{Rohit Pandey}, \bibinfo{person}{Sergio Orts-Escolano}, \bibinfo{person}{Chloe Legendre}, \bibinfo{person}{Christian Haene}, \bibinfo{person}{Sofien Bouaziz}, \bibinfo{person}{Christoph Rhemann}, \bibinfo{person}{Paul~E Debevec}, {and} \bibinfo{person}{Sean~Ryan Fanello}.} \bibinfo{year}{2021}\natexlab{}.
\newblock \showarticletitle{Total relighting: learning to relight portraits for background replacement.}
\newblock \bibinfo{journal}{\emph{ACM Transactions on Graphics}} \bibinfo{volume}{40}, \bibinfo{number}{4} (\bibinfo{year}{2021}), \bibinfo{pages}{43--1}.
\newblock


\bibitem[Ponglertnapakorn et~al\mbox{.}(2023)]%
        {ponglertnapakorn2023difareli}
\bibfield{author}{\bibinfo{person}{Puntawat Ponglertnapakorn}, \bibinfo{person}{Nontawat Tritrong}, {and} \bibinfo{person}{Supasorn Suwajanakorn}.} \bibinfo{year}{2023}\natexlab{}.
\newblock \showarticletitle{DiFaReli: Diffusion face relighting}. In \bibinfo{booktitle}{\emph{Proceedings of the IEEE/CVF International Conference on Computer Vision}}. \bibinfo{pages}{22646--22657}.
\newblock


\bibitem[Prince(2011)]%
        {prince2011digital}
\bibfield{author}{\bibinfo{person}{Stephen Prince}.} \bibinfo{year}{2011}\natexlab{}.
\newblock \bibinfo{booktitle}{\emph{Digital visual effects in cinema: The seduction of reality}}.
\newblock \bibinfo{publisher}{Rutgers University Press}.
\newblock


\bibitem[Rao et~al\mbox{.}(2024)]%
        {rao2024scriptviz}
\bibfield{author}{\bibinfo{person}{Anyi Rao}, \bibinfo{person}{Jean-Pe{\"\i}c Chou}, {and} \bibinfo{person}{Maneesh Agrawala}.} \bibinfo{year}{2024}\natexlab{}.
\newblock \showarticletitle{ScriptViz: A Visualization Tool to Aid Scriptwriting based on a Large Movie Database}. In \bibinfo{booktitle}{\emph{Proceedings of the 37th Annual ACM Symposium on User Interface Software and Technology}}. \bibinfo{pages}{1--13}.
\newblock


\bibitem[Rao et~al\mbox{.}(2023)]%
        {rao2023dynamic}
\bibfield{author}{\bibinfo{person}{Anyi Rao}, \bibinfo{person}{Xuekun Jiang}, \bibinfo{person}{Yuwei Guo}, \bibinfo{person}{Linning Xu}, \bibinfo{person}{Lei Yang}, \bibinfo{person}{Libiao Jin}, \bibinfo{person}{Dahua Lin}, {and} \bibinfo{person}{Bo Dai}.} \bibinfo{year}{2023}\natexlab{}.
\newblock \showarticletitle{Dynamic storyboard generation in an engine-based virtual environment for video production}.
\newblock In \bibinfo{booktitle}{\emph{ACM SIGGRAPH 2023}}.
\newblock


\bibitem[Razali et~al\mbox{.}(2011)]%
        {razali2011power}
\bibfield{author}{\bibinfo{person}{Nornadiah~Mohd Razali}, \bibinfo{person}{Yap~Bee Wah}, {et~al\mbox{.}}} \bibinfo{year}{2011}\natexlab{}.
\newblock \showarticletitle{Power comparisons of shapiro-wilk, kolmogorov-smirnov, lilliefors and anderson-darling tests}.
\newblock \bibinfo{journal}{\emph{Journal of statistical modeling and analytics}} \bibinfo{volume}{2}, \bibinfo{number}{1} (\bibinfo{year}{2011}), \bibinfo{pages}{21--33}.
\newblock


\bibitem[Rusu and Rusu(2024)]%
        {rusu2024script}
\bibfield{author}{\bibinfo{person}{Adrian Rusu} {and} \bibinfo{person}{Amalia Rusu}.} \bibinfo{year}{2024}\natexlab{}.
\newblock \showarticletitle{Script-to-Storyboard-to-Story Reel Framework}. In \bibinfo{booktitle}{\emph{2024 28th International Conference Information Visualisation (IV)}}. IEEE, \bibinfo{pages}{350--355}.
\newblock


\bibitem[Samy et~al\mbox{.}(2025)]%
        {samy2025revolutionizing}
\bibfield{author}{\bibinfo{person}{Tassneam~M Samy}, \bibinfo{person}{Beshoy~I Asham}, \bibinfo{person}{Salwa~O Slim}, {and} \bibinfo{person}{Amr~A Abohany}.} \bibinfo{year}{2025}\natexlab{}.
\newblock \showarticletitle{Revolutionizing online shopping with FITMI: a realistic virtual try-on solution}.
\newblock \bibinfo{journal}{\emph{Neural Computing and Applications}} (\bibinfo{year}{2025}), \bibinfo{pages}{1--20}.
\newblock


\bibitem[Sciortino(2020)]%
        {sciortino2020pre}
\bibfield{author}{\bibinfo{person}{Christine Sciortino}.} \bibinfo{year}{2020}\natexlab{}.
\newblock \showarticletitle{Pre-Production}.
\newblock In \bibinfo{booktitle}{\emph{Makeup Artistry for Film and Television}}. \bibinfo{publisher}{Routledge}, \bibinfo{pages}{18--59}.
\newblock


\bibitem[Shi et~al\mbox{.}(2020)]%
        {shi2020emog}
\bibfield{author}{\bibinfo{person}{Yang Shi}, \bibinfo{person}{Nan Cao}, \bibinfo{person}{Xiaojuan Ma}, \bibinfo{person}{Siji Chen}, {and} \bibinfo{person}{Pei Liu}.} \bibinfo{year}{2020}\natexlab{}.
\newblock \showarticletitle{EmoG: supporting the sketching of emotional expressions for storyboarding}. In \bibinfo{booktitle}{\emph{Proceedings of the 2020 CHI conference on human factors in computing systems}}. \bibinfo{pages}{1--12}.
\newblock


\bibitem[Simon(2012)]%
        {simon2012storyboards}
\bibfield{author}{\bibinfo{person}{Mark Simon}.} \bibinfo{year}{2012}\natexlab{}.
\newblock \bibinfo{booktitle}{\emph{Storyboards: motion in art}}.
\newblock \bibinfo{publisher}{Routledge}.
\newblock


\bibitem[Tan(2018)]%
        {tan2018psychology}
\bibfield{author}{\bibinfo{person}{Ed~S Tan}.} \bibinfo{year}{2018}\natexlab{}.
\newblock \showarticletitle{A psychology of the film}.
\newblock \bibinfo{journal}{\emph{Palgrave Communications}} \bibinfo{volume}{4}, \bibinfo{number}{1} (\bibinfo{year}{2018}).
\newblock


\bibitem[Wang et~al\mbox{.}(2024)]%
        {wang2024roomdreaming}
\bibfield{author}{\bibinfo{person}{Shun-Yu Wang}, \bibinfo{person}{Wei-Chung Su}, \bibinfo{person}{Serena Chen}, \bibinfo{person}{Ching-Yi Tsai}, \bibinfo{person}{Marta Misztal}, \bibinfo{person}{Katherine~M Cheng}, \bibinfo{person}{Alwena Lin}, \bibinfo{person}{Yu Chen}, {and} \bibinfo{person}{Mike~Y Chen}.} \bibinfo{year}{2024}\natexlab{}.
\newblock \showarticletitle{Roomdreaming: Generative-AI approach to facilitating iterative, preliminary interior design exploration}. In \bibinfo{booktitle}{\emph{Proceedings of the 2024 CHI Conference on Human Factors in Computing Systems}}. \bibinfo{pages}{1--20}.
\newblock


\bibitem[Wei et~al\mbox{.}(2024a)]%
        {wei2024hearing}
\bibfield{author}{\bibinfo{person}{Zheng Wei}, \bibinfo{person}{Yuzheng Chen}, \bibinfo{person}{Wai Tong}, \bibinfo{person}{Xuan Zong}, \bibinfo{person}{Huamin Qu}, \bibinfo{person}{Xian Xu}, {and} \bibinfo{person}{Lik-Hang Lee}.} \bibinfo{year}{2024}\natexlab{a}.
\newblock \showarticletitle{Hearing the Moment with MetaEcho! From Physical to Virtual in Synchronized Sound Recording}. In \bibinfo{booktitle}{\emph{Proceedings of the 32nd ACM International Conference on Multimedia}}. \bibinfo{pages}{6520--6529}.
\newblock


\bibitem[Wei et~al\mbox{.}(2024b)]%
        {wei2024multi}
\bibfield{author}{\bibinfo{person}{Zheng Wei}, \bibinfo{person}{Shan Jin}, \bibinfo{person}{Wai Tong}, \bibinfo{person}{David Kei~Man Yip}, \bibinfo{person}{Pan Hui}, {and} \bibinfo{person}{Xian Xu}.} \bibinfo{year}{2024}\natexlab{b}.
\newblock \showarticletitle{Multi-Role VR Training System for Film Production: Enhancing Collaboration with MetaCrew}.
\newblock In \bibinfo{booktitle}{\emph{ACM SIGGRAPH 2024 Posters}}. \bibinfo{pages}{1--2}.
\newblock


\bibitem[Wei et~al\mbox{.}(2025)]%
        {wei2025illuminating}
\bibfield{author}{\bibinfo{person}{Zheng Wei}, \bibinfo{person}{Jia Sun}, \bibinfo{person}{Junxiang Liao}, \bibinfo{person}{Lik-Hang Lee}, \bibinfo{person}{Chan~In Sio}, \bibinfo{person}{Pan Hui}, \bibinfo{person}{Huamin Qu}, \bibinfo{person}{Wai Tong}, {and} \bibinfo{person}{Xian Xu}.} \bibinfo{year}{2025}\natexlab{}.
\newblock \showarticletitle{Illuminating the Scene: How Virtual Environments and Learning Modes Shape Film Lighting Mastery in Virtual Reality}.
\newblock \bibinfo{journal}{\emph{IEEE Transactions on Visualization and Computer Graphics}} (\bibinfo{year}{2025}).
\newblock


\bibitem[Wei et~al\mbox{.}(2023)]%
        {wei2023feeling}
\bibfield{author}{\bibinfo{person}{Zheng Wei}, \bibinfo{person}{Xian Xu}, \bibinfo{person}{Lik-Hang Lee}, \bibinfo{person}{Wai Tong}, \bibinfo{person}{Huamin Qu}, {and} \bibinfo{person}{Pan Hui}.} \bibinfo{year}{2023}\natexlab{}.
\newblock \showarticletitle{Feeling Present! From Physical to Virtual Cinematography Lighting Education with Metashadow}. In \bibinfo{booktitle}{\emph{Proceedings of the 31st ACM International Conference on Multimedia}}. \bibinfo{pages}{1127--1136}.
\newblock


\bibitem[Wu et~al\mbox{.}(2024a)]%
        {wu2024semi}
\bibfield{author}{\bibinfo{person}{Hongtao Wu}, \bibinfo{person}{Yijun Yang}, \bibinfo{person}{Angelica~I Aviles-Rivero}, \bibinfo{person}{Jingjing Ren}, \bibinfo{person}{Sixiang Chen}, \bibinfo{person}{Haoyu Chen}, {and} \bibinfo{person}{Lei Zhu}.} \bibinfo{year}{2024}\natexlab{a}.
\newblock \showarticletitle{Semi-supervised Video Desnowing Network via Temporal Decoupling Experts and Distribution-Driven Contrastive Regularization}. In \bibinfo{booktitle}{\emph{European Conference on Computer Vision}}. Springer, \bibinfo{pages}{70--89}.
\newblock


\bibitem[Wu et~al\mbox{.}(2023)]%
        {wu2023mask}
\bibfield{author}{\bibinfo{person}{Hongtao Wu}, \bibinfo{person}{Yijun Yang}, \bibinfo{person}{Haoyu Chen}, \bibinfo{person}{Jingjing Ren}, {and} \bibinfo{person}{Lei Zhu}.} \bibinfo{year}{2023}\natexlab{}.
\newblock \showarticletitle{Mask-guided progressive network for joint raindrop and rain streak removal in videos}. In \bibinfo{booktitle}{\emph{Proceedings of the 31st ACM International Conference on Multimedia}}. \bibinfo{pages}{7216--7225}.
\newblock


\bibitem[Wu et~al\mbox{.}(2024b)]%
        {wu2024rainmamba}
\bibfield{author}{\bibinfo{person}{Hongtao Wu}, \bibinfo{person}{Yijun Yang}, \bibinfo{person}{Huihui Xu}, \bibinfo{person}{Weiming Wang}, \bibinfo{person}{Jinni Zhou}, {and} \bibinfo{person}{Lei Zhu}.} \bibinfo{year}{2024}\natexlab{b}.
\newblock \showarticletitle{Rainmamba: Enhanced locality learning with state space models for video deraining}. In \bibinfo{booktitle}{\emph{Proceedings of the 32nd ACM International Conference on Multimedia}}. \bibinfo{pages}{7881--7890}.
\newblock


\bibitem[Xie et~al\mbox{.}(2023)]%
        {xie2023wakey}
\bibfield{author}{\bibinfo{person}{Liwenhan Xie}, \bibinfo{person}{Zhaoyu Zhou}, \bibinfo{person}{Kerun Yu}, \bibinfo{person}{Yun Wang}, \bibinfo{person}{Huamin Qu}, {and} \bibinfo{person}{Siming Chen}.} \bibinfo{year}{2023}\natexlab{}.
\newblock \showarticletitle{Wakey-wakey: Animate text by mimicking characters in a gif}. In \bibinfo{booktitle}{\emph{Proceedings of the 36th Annual ACM Symposium on User Interface Software and Technology}}. \bibinfo{pages}{1--14}.
\newblock


\bibitem[Xu et~al\mbox{.}(2020)]%
        {xu2020megatron}
\bibfield{author}{\bibinfo{person}{Peng Xu}, \bibinfo{person}{Mostofa Patwary}, \bibinfo{person}{Mohammad Shoeybi}, \bibinfo{person}{Raul Puri}, \bibinfo{person}{Pascale Fung}, \bibinfo{person}{Anima Anandkumar}, {and} \bibinfo{person}{Bryan Catanzaro}.} \bibinfo{year}{2020}\natexlab{}.
\newblock \showarticletitle{MEGATRON-CNTRL: Controllable story generation with external knowledge using large-scale language models}.
\newblock \bibinfo{journal}{\emph{arXiv preprint arXiv:2010.00840}} (\bibinfo{year}{2020}).
\newblock


\bibitem[Xu et~al\mbox{.}(2024b)]%
        {xu2024transforming}
\bibfield{author}{\bibinfo{person}{Xian Xu}, \bibinfo{person}{Wai Tong}, \bibinfo{person}{Zheng Wei}, \bibinfo{person}{Meng Xia}, \bibinfo{person}{Lik-Hang Lee}, {and} \bibinfo{person}{Huamin Qu}.} \bibinfo{year}{2024}\natexlab{b}.
\newblock \showarticletitle{Transforming cinematography lighting education in the metaverse}.
\newblock \bibinfo{journal}{\emph{Visual Informatics}} (\bibinfo{year}{2024}).
\newblock


\bibitem[Xu et~al\mbox{.}(2024a)]%
        {xu2024skipwriter}
\bibfield{author}{\bibinfo{person}{Zheer Xu}, \bibinfo{person}{Shanqing Cai}, \bibinfo{person}{Mukund Varma~T}, \bibinfo{person}{Subhashini Venugopalan}, {and} \bibinfo{person}{Shumin Zhai}.} \bibinfo{year}{2024}\natexlab{a}.
\newblock \showarticletitle{SkipWriter: LLM-Powered Abbreviated Writing on Tablets}. In \bibinfo{booktitle}{\emph{Proceedings of the 37th Annual ACM Symposium on User Interface Software and Technology}}. \bibinfo{pages}{1--13}.
\newblock


\bibitem[Yang et~al\mbox{.}(2024a)]%
        {yang2024emogen}
\bibfield{author}{\bibinfo{person}{Jingyuan Yang}, \bibinfo{person}{Jiawei Feng}, {and} \bibinfo{person}{Hui Huang}.} \bibinfo{year}{2024}\natexlab{a}.
\newblock \showarticletitle{EmoGen: Emotional image content generation with text-to-image diffusion models}. In \bibinfo{booktitle}{\emph{Proceedings of the IEEE/CVF Conference on Computer Vision and Pattern Recognition}}. \bibinfo{pages}{6358--6368}.
\newblock


\bibitem[Yang et~al\mbox{.}(2024b)]%
        {yang2024genuine}
\bibfield{author}{\bibinfo{person}{Yijun Yang}, \bibinfo{person}{Hongtao Wu}, \bibinfo{person}{Angelica~I Aviles-Rivero}, \bibinfo{person}{Yulun Zhang}, \bibinfo{person}{Jing Qin}, {and} \bibinfo{person}{Lei Zhu}.} \bibinfo{year}{2024}\natexlab{b}.
\newblock \showarticletitle{Genuine knowledge from practice: Diffusion test-time adaptation for video adverse weather removal}. In \bibinfo{booktitle}{\emph{2024 IEEE/CVF Conference on Computer Vision and Pattern Recognition (CVPR)}}. IEEE, \bibinfo{pages}{25606--25616}.
\newblock


\bibitem[Yin et~al\mbox{.}(2023)]%
        {yin2023cle}
\bibfield{author}{\bibinfo{person}{Yuyang Yin}, \bibinfo{person}{Dejia Xu}, \bibinfo{person}{Chuangchuang Tan}, \bibinfo{person}{Ping Liu}, \bibinfo{person}{Yao Zhao}, {and} \bibinfo{person}{Yunchao Wei}.} \bibinfo{year}{2023}\natexlab{}.
\newblock \showarticletitle{Cle diffusion: Controllable light enhancement diffusion model}. In \bibinfo{booktitle}{\emph{Proceedings of the 31st ACM International Conference on Multimedia}}. \bibinfo{pages}{8145--8156}.
\newblock


\bibitem[Yuan et~al\mbox{.}(2022)]%
        {yuan2022wordcraft}
\bibfield{author}{\bibinfo{person}{Ann Yuan}, \bibinfo{person}{Andy Coenen}, \bibinfo{person}{Emily Reif}, {and} \bibinfo{person}{Daphne Ippolito}.} \bibinfo{year}{2022}\natexlab{}.
\newblock \showarticletitle{Wordcraft: story writing with large language models}. In \bibinfo{booktitle}{\emph{Proceedings of the 27th International Conference on Intelligent User Interfaces}}. \bibinfo{pages}{841--852}.
\newblock


\bibitem[Yue et~al\mbox{.}(2023)]%
        {yue2023dif}
\bibfield{author}{\bibinfo{person}{Jun Yue}, \bibinfo{person}{Leyuan Fang}, \bibinfo{person}{Shaobo Xia}, \bibinfo{person}{Yue Deng}, {and} \bibinfo{person}{Jiayi Ma}.} \bibinfo{year}{2023}\natexlab{}.
\newblock \showarticletitle{Dif-fusion: Toward high color fidelity in infrared and visible image fusion with diffusion models}.
\newblock \bibinfo{journal}{\emph{IEEE Transactions on Image Processing}}  \bibinfo{volume}{32} (\bibinfo{year}{2023}), \bibinfo{pages}{5705--5720}.
\newblock


\bibitem[Zeng et~al\mbox{.}(2024)]%
        {zeng2024dilightnet}
\bibfield{author}{\bibinfo{person}{Chong Zeng}, \bibinfo{person}{Yue Dong}, \bibinfo{person}{Pieter Peers}, \bibinfo{person}{Youkang Kong}, \bibinfo{person}{Hongzhi Wu}, {and} \bibinfo{person}{Xin Tong}.} \bibinfo{year}{2024}\natexlab{}.
\newblock \showarticletitle{Dilightnet: Fine-grained lighting control for diffusion-based image generation}. In \bibinfo{booktitle}{\emph{ACM SIGGRAPH 2024 Conference Papers}}. \bibinfo{pages}{1--12}.
\newblock


\bibitem[Zhang et~al\mbox{.}(2024a)]%
        {zhang2024brush}
\bibfield{author}{\bibinfo{person}{Lingjun Zhang}, \bibinfo{person}{Xinyuan Chen}, \bibinfo{person}{Yaohui Wang}, \bibinfo{person}{Yue Lu}, {and} \bibinfo{person}{Yu Qiao}.} \bibinfo{year}{2024}\natexlab{a}.
\newblock \showarticletitle{Brush your text: Synthesize any scene text on images via diffusion model}. In \bibinfo{booktitle}{\emph{Proceedings of the AAAI Conference on Artificial Intelligence}}, Vol.~\bibinfo{volume}{38}. \bibinfo{pages}{7215--7223}.
\newblock


\bibitem[Zhang et~al\mbox{.}(2023)]%
        {zhang2023adding}
\bibfield{author}{\bibinfo{person}{Lvmin Zhang}, \bibinfo{person}{Anyi Rao}, {and} \bibinfo{person}{Maneesh Agrawala}.} \bibinfo{year}{2023}\natexlab{}.
\newblock \showarticletitle{Adding conditional control to text-to-image diffusion models}. In \bibinfo{booktitle}{\emph{Proceedings of the IEEE/CVF international conference on computer vision}}. \bibinfo{pages}{3836--3847}.
\newblock


\bibitem[Zhang et~al\mbox{.}(2024b)]%
        {zhang2024ic}
\bibfield{author}{\bibinfo{person}{Lvmin Zhang}, \bibinfo{person}{Anyi Rao}, {and} \bibinfo{person}{Maneesh Agrawala}.} \bibinfo{year}{2024}\natexlab{b}.
\newblock \showarticletitle{Ic-light github page}.
\newblock  (\bibinfo{year}{2024}).
\newblock


\bibitem[Zhang et~al\mbox{.}(2025)]%
        {zhang2025generative}
\bibfield{author}{\bibinfo{person}{Ruihan Zhang}, \bibinfo{person}{Borou Yu}, \bibinfo{person}{Jiajian Min}, \bibinfo{person}{Yetong Xin}, \bibinfo{person}{Zheng Wei}, \bibinfo{person}{Juncheng~Nemo Shi}, \bibinfo{person}{Mingzhen Huang}, \bibinfo{person}{Xianghao Kong}, \bibinfo{person}{Nix~Liu Xin}, \bibinfo{person}{Shanshan Jiang}, {et~al\mbox{.}}} \bibinfo{year}{2025}\natexlab{}.
\newblock \showarticletitle{Generative AI for Film Creation: A Survey of Recent Advances}. In \bibinfo{booktitle}{\emph{Proceedings of the Computer Vision and Pattern Recognition Conference}}. \bibinfo{pages}{6267--6279}.
\newblock


\bibitem[Zhang and Liu(2024)]%
        {zhang2024unlocking}
\bibfield{author}{\bibinfo{person}{Yanbo Zhang} {and} \bibinfo{person}{Chuanlan Liu}.} \bibinfo{year}{2024}\natexlab{}.
\newblock \showarticletitle{Unlocking the potential of artificial intelligence in fashion design and e-commerce applications: The case of Midjourney}.
\newblock \bibinfo{journal}{\emph{Journal of Theoretical and Applied Electronic Commerce Research}} \bibinfo{volume}{19}, \bibinfo{number}{1} (\bibinfo{year}{2024}), \bibinfo{pages}{654--670}.
\newblock


\bibitem[Zhao(2023)]%
        {zhao2023leveraging}
\bibfield{author}{\bibinfo{person}{Xin Zhao}.} \bibinfo{year}{2023}\natexlab{}.
\newblock \showarticletitle{Leveraging artificial intelligence (AI) technology for English writing: Introducing wordtune as a digital writing assistant for EFL writers}.
\newblock \bibinfo{journal}{\emph{RELC Journal}} \bibinfo{volume}{54}, \bibinfo{number}{3} (\bibinfo{year}{2023}), \bibinfo{pages}{890--894}.
\newblock


\bibitem[Zhu et~al\mbox{.}(2023)]%
        {zhu2023minigpt}
\bibfield{author}{\bibinfo{person}{Deyao Zhu}, \bibinfo{person}{Jun Chen}, \bibinfo{person}{Xiaoqian Shen}, \bibinfo{person}{Xiang Li}, {and} \bibinfo{person}{Mohamed Elhoseiny}.} \bibinfo{year}{2023}\natexlab{}.
\newblock \showarticletitle{Minigpt-4: Enhancing vision-language understanding with advanced large language models}.
\newblock \bibinfo{journal}{\emph{arXiv preprint arXiv:2304.10592}} (\bibinfo{year}{2023}).
\newblock


\end{thebibliography}

\end{document}